\documentclass[fleqn,usenatbib]{mnras}

\usepackage{newtxtext,newtxmath}

\usepackage[T1]{fontenc}

\DeclareRobustCommand{\VAN}[3]{#2}
\let\VANthebibliography\thebibliography
\def\thebibliography{\DeclareRobustCommand{\VAN}[3]{##3}\VANthebibliography}

\usepackage{graphicx}	
\usepackage{amsmath}	
\usepackage{amssymb}	

\newcommand{\HI}{\ion{H}{i}\,} 	
\newcommand{\nside}{\texttt{nside}\,}  

\title[GMCA foreground cleaning for 21cm IM experiments]{Recovery of 21\,cm intensity maps with  sparse component separation}

\author[I. P. Carucci et al.]{
	Isabella P. Carucci\thanks{E-mail: isabella.carucci@cea.fr},
	Melis O. Irfan
	and J\'er\^ome Bobin
	\\
	AIM, CEA, CNRS, Universit\'e Paris-Saclay,  Universit\'e  Paris Diderot, Sorbonne Paris Cit\'e, F-91191 Gif-sur-Yvette, France
}

\date{Accepted XXX. Received YYY; in original form ZZZ}

\pubyear{2020}

\begin{document}
\label{firstpage}
\pagerange{\pageref{firstpage}--\pageref{lastpage}}
\maketitle

\begin{abstract}
21\,cm intensity mapping has emerged as a promising technique to map the large-scale structure of the Universe. However, the presence of foregrounds with amplitudes orders of magnitude larger than the cosmological signal constitutes a critical challenge. Here, we test the sparsity-based algorithm Generalised Morphological Component Analysis (GMCA)  as a blind component separation technique for this class of experiments. We test the GMCA performance against realistic full-sky mock temperature maps that include, besides astrophysical foregrounds, also a fraction of the polarized part of the signal leaked into the unpolarized one, a very troublesome foreground to subtract, usually referred to as polarization leakage. To our knowledge, this is the first time the removal of such component is performed with no prior assumption. We assess the success of the cleaning by comparing the true and recovered power spectra, in the angular and radial directions. In the best scenario looked at, GMCA is able to recover the input angular (radial) power spectrum with an average bias of  $\sim 5\%$ for $\ell>25$ ($20 - 30 \%$ for $k_{\parallel} \gtrsim 0.02 \,h^{-1}$Mpc), in the presence of polarization leakage. Our results are robust also when up to $40\%$ of channels are missing, mimicking a Radio Frequency Interference (RFI) flagging of the data. Having quantified the notable effect of polarisation leakage on our results, in perspective we advocate the use of more realistic simulations when testing 21 cm intensity mapping capabilities.
\end{abstract}

\begin{keywords}
methods: data analysis -- methods: statistical -- cosmology: observations -- large-scale structure of universe -- radio lines: galaxies -- radio lines: ISM.
\end{keywords}



\section{Introduction}

If we would ask a large-scale structure scientist about her ideal survey, she would request big cosmological volumes and great redshift resolution. Both things are hard to achieve at the same time. For instance, if we consider galaxy surveys, those are either photometric (big volumes but also big redshift errors) or spectroscopic (accurate redshifts but small volumes). This motivates the development of 21\,cm intensity mapping experiments that can ensure both advantages.

Indeed, the 21\,cm -- alternatively, the frequency $\nu_{\rm 21cm} = 1420$ MHz -- line is emitted by the hyperfine transition of neutral hydrogen, \HI. Being spectrally isolated, we are confident we are observing a \HI cloud at redshift $z$ when detecting a signal at frequency
$ \nu = \nu_{\rm 21cm}/(1+z) $. Hydrogen is the most abundant baryonic component of the Universe, however, its 21\,cm line is weak and long integration times are necessary to detect galaxies beyond $z\gtrsim0.1$ \citep[e.g.][]{fernandez2016}. To overcome this, we can use the intensity mapping technique: we drop the idea of resolving individual galaxies and instead collect all their integrated emission, scanning the sky fast and economically. This way, we tomographically assemble temperature maps in 21\,cm of the Universe, effectively mapping the cosmic web in three dimensions \citep{Battye2004,Chang2008,Loeb2008}.

However, since its first application in cross-correlation with galaxies by \citet{Chang2010} with Green Bank Telescope (GBT) data, 21cm intensity mapping has proven to be hard to be performed. There have been updates with GBT data  \citep{Masui2013,Switzer2013,wolz2017} and more recently also with the Parkes radio telescope, although still in cross-correlation with galaxies \citep{anderson2018}. We still miss a truly independent detection.

The main challenge for these experiments is constituted by contaminants: foregrounds of astrophysical origin -- that are orders of magnitude more intense than the sought-after signal -- and those originated by instrumental issues, as systematics and calibration-driven effects; the latter can also mix the modes (of foregrounds and signal), making even more challenging the component separation \citep{Switzer2015}. The discussion about more adapted and optimised cleaning methods motivates this paper.

Many of the foreground cleaning methods tested in the literature make use of the expected smoothness in frequency of the astrophysical foregrounds. Some of them parametrize the foregrounds in order to separate them \citep{Ansari2012,Shaw2014}, others do not assume a specific model for the foregrounds and are said to be blind --  Principal Component Analysis \citep{alonso2015,bigot-sazy2015}; Independent Component Analysis \citep{wolz2014,Zhang2016,Cunnington2019}; inverse variance \citep{Liu2011}; quadratic estimators \citep{Switzer2015}; Generalized Needlet Internal Linear Combination \citep{olivari2016}. Up to now, in real data analysis, only blind methods have been employed and proven to be suitable for the foreground cleaning task \citep{Masui2013,Switzer2013,wolz2017,anderson2018}.

All the methods and works mentioned above succeed at cleaning the maps with different levels of accuracy. However, none of them include and search for components that are not smooth in frequency. In this paper we upgrade the degree of complexity of the simulated data we want to clean, including a non-smooth component that we expect to manifest in these observations: polarization leakage, a fraction of the polarized part of the signal that spills into the total intensity one. To our knowledge, this is the first time the removal of such kind of contaminant is attempted assuming no prior knowledge about it. We do so using the GMCA algorithm \citep[Generalised Morphological Component Analysis,][]{gmca2007}. It is also the first time GMCA is adapted and tested as a blind source separation method for $z<6$ \HI intensity mapping data. Different versions of GMCA have already been applied to various observational data-sets, as Cosmic Microwave Background data \citep[e.g.][]{gmcaCMB}, 21\,cm interferometric data in the epoch-of-reionization (EoR) context \citep{patil2017} and X-ray images of Supernova remnants \citep{Picquenot2019}. 

A 21\,cm intensity mapping survey can be performed either in single-dish mode (one or more single-dish antennas used as a set of autocorrelators) or in standard interferometry \citep{bull2015}; current and planned surveys exist for both regimes. Here we choose to focus on a survey like MeerKLASS \citep{MeerKLASS}, a proposed single-dish survey with the MeerKAT radio telescope. Nevertheless, the results of this paper could be extended to other instrumental configurations (as for example BINGO\footnote{\url{http://www.bingotelescope.org/en/}} and FAST\footnote{\url{https://fast.bao.ac.cn}}) and also in interferometry as we will point out.

The paper is organized as follows. In Section~\ref{sec:GMCA} we formalise the problem we are tackling and we present the GMCA assumptions and method. In Section~\ref{sec:sims} we describe the simulation we use for testing GMCA.  In Section~\ref{sec:pipeline}, we present how we apply GMCA on the simulated data and how we evaluate the outcome of the foreground removal. In Section~\ref{sec:results} we describe and discuss the obtained results. Finally, we summarise our work in Section~\ref{sec:conclusions}.

\section{Source separation formalism}
\label{sec:GMCA}

\subsection{The 21\,cm intensity mapping context}

An intensity mapping survey scans the sky and for each channel of frequency $\nu$ compiles a map of the total brightness temperature $T$. For each given position on the sky (each pixel $p$) $T$ is the sum of the cosmological 21\,cm signal from \HI, of the foregrounds and of the instrumental noise:
\begin{equation} \label{eq:Tmap}
    T(\nu,p) = T_{C} (\nu,p) + T_{\rm F} (\nu,p) + T_{N}(\nu,p)\,.
\end{equation}
In the source separation process, we think of the foreground contribution $T_{\rm F}$ as a sum of $n_{\rm s}$ sources modulated by a frequency-dependent amplitude, i.e. for each map at $\nu$:
\begin{equation} \label{eq:TF}
    T_{\rm F}(\nu,p) = \sum_{i=1}^{n_{\rm s}} A_i(\nu) S_i(p) \,.
\end{equation}
We compress all maps in a data-cube \textbf{X}, i.e. a  $n_{\rm pix} \times n_{\rm ch}$ matrix with $n_{\rm pix}$ the number of pixels in each map and $n_{\rm ch}$ the number of maps (channels). We merge equations (\ref{eq:Tmap}) and (\ref{eq:TF}) and we can write in matrix form:
\begin{equation} \label{eq:master}
    \textbf{X} = \textbf{A}\textbf{S} + \textbf{C} + \textbf{N}\,,
\end{equation}
where \textbf{A} is the mixing matrix governing the contribution of the $n_{\rm s}$ sources \textbf{S} in the resulting signal, up to the cosmological signal \textbf{C} and the noise contribution \textbf{N}. It follows that \textbf{A} has  $n_{\rm s} \times n_{\rm ch}$ while \textbf{S} has $n_{\rm pix} \times n_{\rm s}$ dimensions.

We recall that the cosmological 21\,cm signal is (i) highly outweighed by the foregrounds and (ii) uncorrelated in frequency. This implies that the cosmological signal \textbf{C} is inherently coupled to the instrumental noise component \textbf{N}. The problem of foreground removal reduces to estimate the foreground driven \textbf{A}\textbf{S} so that \textbf{X} - \textbf{A}\textbf{S} is as accurate as possible at predicting the cosmological \HI brightness temperature field, taking into account the instrumental noise contribution.

\subsection{Generalised Morphological Component Analysis}

GMCA is a blind source separation algorithm that relies on the morphological features that compose the sought-after components. To that purpose, such components are assumed to admit a sparse distribution in an adapted signal representation ({\it e.g.} Fourier, wavelets, to only name two).
A source is sparse when most of its coefficients are zero, thus sparse sources are easier to disentangle as their signatures are uncorrelated. A classic example is Fourier-space for periodic signals: they can be described by few coefficients. The sparsity assumption is essential as it allows to dramatically improve the contrast between distinct components, which ease the separation process.

For the science case of this paper, we make use of the Starlet wavelet dictionary \citep{Starck2007}, that has proven to be well adapted for an efficient sparse description of galactic diffuse emissions and astrophysical images in general \citep[e.g.][]{Floer2014,Joseph2016,Offringa2017,Irfan2018}. 

Once we wavelet-transform  \textbf{X} to \textbf{X}$^{\rm wt}$, GMCA promotes sparsity in the sources \textbf{S}$^{\rm wt}$ in wavelet-base by solving iteratively the following optimisation problem: 
\begin{equation}
\label{eq:gmca}
\{\mathbf{\tilde{A}}, \mathbf{\tilde{S}}\} = \min_{\textbf{A}, \textbf{S}^{\rm wt}}  \sum_{i=1} ^{n_{\rm s}}  \lambda_i \left\lvert \left\lvert \textbf{S}^{\rm wt}_{i} \right\rvert \right\rvert_1+ \left\lvert \left\lvert \textbf{X}^{\rm wt} - \textbf{A} \textbf{S}^{\rm wt}  \right\rvert \right\rvert_{F}^{2}, 
\end{equation}
where the first term is a sparsity constraint term and the second is a data-fidelity term. Indeed, $\left\lvert \left\lvert  \cdot \right\rvert \right\rvert_{1}$ is the $\ell_1$ norm\footnote{For a pure sparse solution, we could substitute it with the $\ell_0$ norm: $\left\lvert \left\lvert  \mathbf{Y} \right\rvert \right\rvert_{0}$, the number of non-zero entries in $\mathbf{Y}$.} defined by $\left\lvert \left\lvert  \mathbf{Y} \right\rvert \right\rvert_{1} = \sum_{i,j} \left\lvert  \mathbf{Y}_{i,j} \right\rvert$. And $\left\lvert \left\lvert  \cdot \right\rvert \right\rvert_{F}$ the Frobenius norm defined by $\left\lvert \left\lvert  \mathbf{Y} \right\rvert \right\rvert_{F}^2 = {\rm Trace}( \mathbf{YY}^T )$.
In particular, $\lambda_i$ are regularisation coefficients --  sparsity-thresholds --  essential to provide robustness with respect to the noise of the problem, i.e. in our case the difference in intensity between the foregrounds and the cosmological signal; we first estimate them with the median absolute deviation (MAD) method and progressively decrease towards a final noise-related level.  We refer the reader to \citet{L-GMCA2} for details about the thresholding strategy.

As neither for $\mathbf{A}$ nor $\mathbf{S}$ we use a model, GMCA is said to be a blind method, where the only input needed is the number of components $n_{\rm s}$ it searches  and its assumption is constituted by sparsity.

The algorithm we employ here is openly available at \url{www.cosmostat.org} and demonstration scripts for reproducing the results of this article are available at \url{https://github.com/isab3lla/gmca4im}.

\section{Simulations}
\label{sec:sims}

\begin{table*}
	\centering
	\caption{Schematic descriptions of the components of the simulated maps.}
	\label{tab:sim_components}
	\begin{tabular}{lll} 
		\hline
		Component & Method / template & Parameters\\
		\hline
		Cosmological  &  lognormal approximation \citep[\texttt{CRIME},][]{alonso2014_CRIME} & $b_{\HI}(z) = 0.3 (1+z) + 0.6$\,,\\
		\,\, 21\,cm signal  & & $\Omega_{\HI}(z) = 4(1+z)^{0.6}10^{-4}$ \\ \\
		Galactic synchrotron & Planck Legacy Archive - FFP10 + & spatially varying spectral index ($\beta_s \sim -3$)\\
		 & \, high resolution padding as in equation~(\ref{eq:pad}) &  \\ \\
		Galactic free-free & Planck Legacy Archive - FFP10 & constant spectral index  $\beta_s = -2.13$\\ \\
		Extra-galactic & empirical model by \citet{battye2013},  & source flux threshold  $S_0=1$ Jy, \,\,background $T \sim \nu^\alpha$ \\
		\,  point sources & \, Poisson and clustering contributions as in \citet{olivari2018} &  ($\alpha$ from a Gaussian distr. with $\alpha_0 = -2.7$, $\sigma = 0.2$)\\ \\
		Polarization leakage & galactic synchrotron polarization + & fraction of leaked polarization $\epsilon=0.5\%$ ,  \\
		   & \, rotation measurement \citep[\texttt{CRIME},][]{alonso2014_CRIME}&  Faraday-space corr. length $\xi_\Psi = 0.5$ rad m$^{-2}$ \\ \\
		Telescope beam & Gaussian smoothing & frequency-dependent $\theta_{\rm FWHM}$, see equation~(\ref{eq:theta}) \\ \\
		Instrumental noise & white,  frequency-dependent $\sigma_N$ as in equation~(\ref{eq:sigmaN}) & see Table \ref{tab:instrument} \\
		\hline
	\end{tabular}
\end{table*}

In this Section we describe the simulated data we test the foreground removal technique on. To reproduce a truthful sky at frequencies of $900 - 1400$ MHz, we sum together different components: (i) the 21cm cosmological signal, for which we use the lognormal approximation as proposed by \citet{alonso2014_CRIME}; (ii) astrophysical foregrounds, of galactic origin -- synchrotron and free-free diffuse emissions -- that we estimate using the Planck Sky Model, and of extragalactic origin, for which we adopt the model by \citet{battye2013}; (iii) lastly, we consider polarization leakage: a systematic known to be critical  in \HI intensity mapping experiments \citep{santos2015}, that we model as in \citet{alonso2014_CRIME}. polarization leakage is difficult to deal with because it is expected to be non-smooth in frequency -- as we will later explicitly show -- and common foreground removal techniques are not aimed at picking components {\it misbehaving} in frequency. Thus, very little has been done in the literature to attempt to remove its contribution from the signal \citep{Shaw2015} and to our knowledge there have been no attempts to remove it blindly, i.e. assuming no prior on its characteristics.

For each frequency and for all components, we generate \texttt{HEALPIX} maps with \nside $=256$, that correspond to $n_{\rm pix}=12\, {\rm \nside}^2$ pixels per map \citep{healpix}. We make this {\it parent} simulation publicly available at \url{http://doi.org/10.5281/zenodo.3991818} \citep{simulation}; all scenarios addressed in this paper can be derived from it.

We then merge all components into sky maps and we mimic survey-specific features: we smooth maps with a frequency-dependent Gaussian filter to mimic the effect of the telescope beam and we finally add white noise to each channel, following standard thermal noise calculations. 
In the next paragraphs, we describe with more detail each of the above steps. Main properties of the simulated components are summarised in Table \ref{tab:sim_components}.

\subsection{Cosmological signal}

After reionization ($z<6$), most \HI in the Universe is stored inside galaxies, where it is dense enough to self-shield against the ionizing power of the cosmic ultra-violet background \citep{Noterdaeme2012,Zafar2013}.
Thus, we can associate \HI to the densest regions of the underlying dark matter field: we approximate the latter by a lognormal realisation \citep{lognormal} and assume \HI is its linear biased tracer. We make use of the \texttt{CRIME}\footnote{\url{http://intensitymapping.physics.ox.ac.uk/CRIME.html}} algorithm, described in \citet{alonso2014_CRIME}. We minimally modify \texttt{CRIME}, as we choose to set a redshift-dependent \HI bias $b_{\HI}(z) = 0.3 (1+z) + 0.6$ in agreement with observations at redshift $z \lesssim 0.8$ \citep{Martin2012,Switzer2013} and to set the overall \HI cosmic abundance to $\Omega_{\HI}(z) = 4(1+z)^{0.6}10^{-4}$, as compiled by \citet{Crighton2015}.

The lognormal realisation has cosmological parameters $\{\Omega_{m}, \Omega_{\Lambda}, \Omega_{b}, h\} = \{0.3, 0.7, 0.049, 0.67\}$, with an initial cube of side 3 Gpc/$h$  divided in $2048^3$ cells. Light-cone effects and redshift-space distortions are included by construction. The original simulation is composed by 400 channels of 1\,MHz of thickness, covering a redshift range of $z \in [0.09 -0.58]$, corresponding to frequencies $\nu \in  [900 - 1300]$ MHz. We later re-bin the simulation as described in  Section~\ref{sec:datacubes} before performing the blind source separation.

The lognormal approximation is appropriate for this study, especially considering that we later smooth the maps with a typical beam of $\approx 1 \deg$, losing the small scale information of the field. The large scale properties displayed by the simulated \HI field roughly match those seen in local Universe \HI galaxy survey \citep{Obuljen2019}, in state-of-the-art hydro-dynamical simulations \citep{paco2018_illustris} and in state-of-the-art galaxy evolution models coupled to N-body simulations \citep{Spinelli2020}.

\subsection{Astrophysical foregrounds}

\begin{figure}
	\includegraphics[width=\columnwidth]{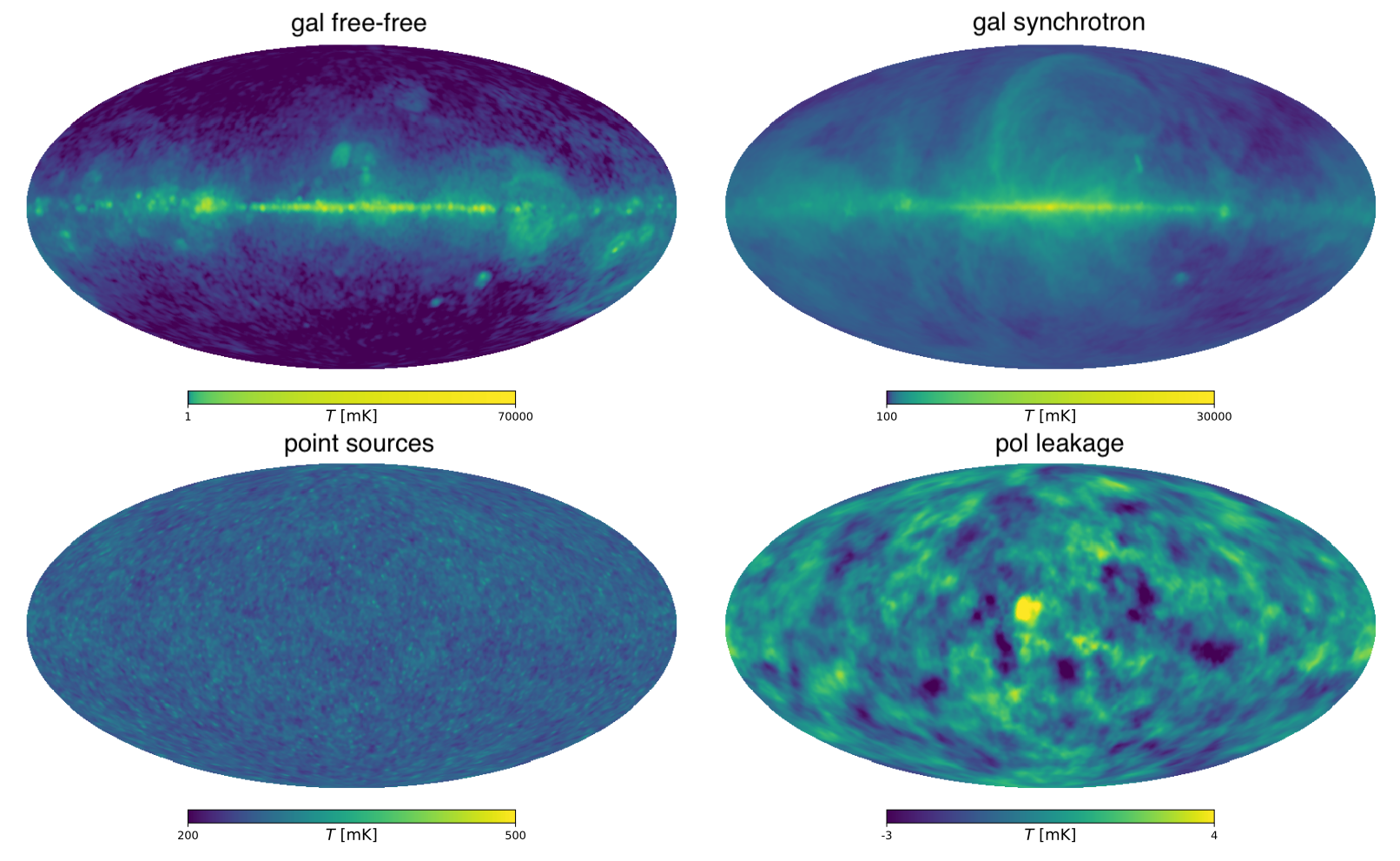}
	\caption{Mollweide projections of the temperature of the contaminant components in the simulation, from top left clockwise: galactic free-free, galactic synchrotron, polarization leakage and point sources. Units are in mK and the colour bar is in logarithmic (linear) scale for the top (bottom) maps. Maps correspond to frequency $\nu =1101$~MHz and have been convolved with the telescope beam as described in the text.}
	\label{fig:molls_sim}
\end{figure}

The astrophysical foregrounds featured in these simulations can be divided into two groups: galactic and extragalactic. For the extragalactic radio sources we implement the empirical model of \citet{battye2013}, who obtain their differential source counts from an empirical fit to numerous 1.4\,GHz source surveys. By integrating these source counts a mean offset temperature, representing the unresolved sources is calculated for 1.4\,GHz. The point sources also contribute a clustering and Poisson component to the overall point source temperature per pixel; these are calculated in angular power space and then converted to pixel space using the \texttt{HEALPIX} \texttt{synfast} routine. Finally, any point sources over 0.01\,Jy are injected at random into the map as fully resolved sources using the map pixel area and the number of sources (N) per steradian with a flux of S \citep{olivari2016}:
\begin{equation}
T_{ps}(1.4\,{\rm{GHz}}, p) = \left( \frac{\lambda^{2}}{2k_{B}} \right) \Omega_{pix}^{-1} \sum_{i=1}^{N} S_{i}
\end{equation}

We assume that sources brighter than 1 Jy have been identified and removed from the data. In order to scale this 1.4\,GHz estimate across our frequency range we use a power law where the spectral index varies, according to a Gaussian distribution, over the sky. For the Gaussian distribution we choose a mean value of -2.7 and a standard deviation of 0.2 \citep{bigot-sazy2015}.

The diffuse galactic foregrounds present in intensity at MHz frequencies are synchrotron and free-free emission. These emissions can both be modelled at each pixel $p$ as power laws with an amplitude $T_{s}$ and spectral index $\beta_{s}$:
\begin{equation}
T(\nu, p) = T_s(p) \left (\frac{\nu}{\nu_{0}}\right)^{\beta_s(\nu,p)}.     
\end{equation}
The Planck Legacy Archive\footnotemark \footnotetext{\url{http://pla.esac.esa.int/pla}} FFP10 simulations provide the synchrotron and free-free all-sky amplitudes and as well as the synchrotron spectral index. We use the FFP10 simulations at 217\,GHz for the free-free and synchrotron amplitudes, at \nside 2048 and degrade and smooth these maps to our desired \nside and resolution using the \texttt{HEALPIX} routines. The synchrotron spectral index map used ($\beta$) is that of \citet{mamd08} and is at a resolution of around 5 degrees. To provide spectral index information at angular scales smaller than 5 degrees we combine the synchrotron spectral map with a map of small scale structure:
\begin{equation}
\beta_{sy} = \beta + \beta_{ss},
\end{equation}
where the small-scale fluctuations ($ \beta_{ss}$), are taken from \citet{santos2005} and adapted to have a smaller amplitude:
\begin{equation} \label{eq:pad}
C_{\ell}^{\beta_{ss}} =  7 \times 10^{-6} \left( \frac{1000}{\ell} \right)^{2.4} \times \left(\frac{\nu_{r}^{2}}{\nu_{1} \nu_{2}}\right)^{2.8} \times {\rm{exp}}\left(\frac{- {\rm{log}}(\nu_{1} / \nu_{2})^{2}}{2 \times 4^{2}} \right),
\end{equation}
where $\nu_{r}$ is 130\,MHz, $\nu_{1}$ is 580\,MHz and $\nu_{2}$ is 1000\,MHz.

The synchrotron spectral index varies across pixels; this is not the case for the free-free spectral index. In alignment with the known range of -2.15 to -2.10 \citep{clive}, we set the value for the free-free spectral index to be -2.13 and keep this constant across the whole sky and over our full frequency range.  

Fig.~\ref{fig:molls_sim} shows the all-sky foreground maps which constitute the astrophysical components of our simulation.

\subsection{A non-smooth contaminant: polarization leakage}

\begin{figure}
	\includegraphics[width=\columnwidth]{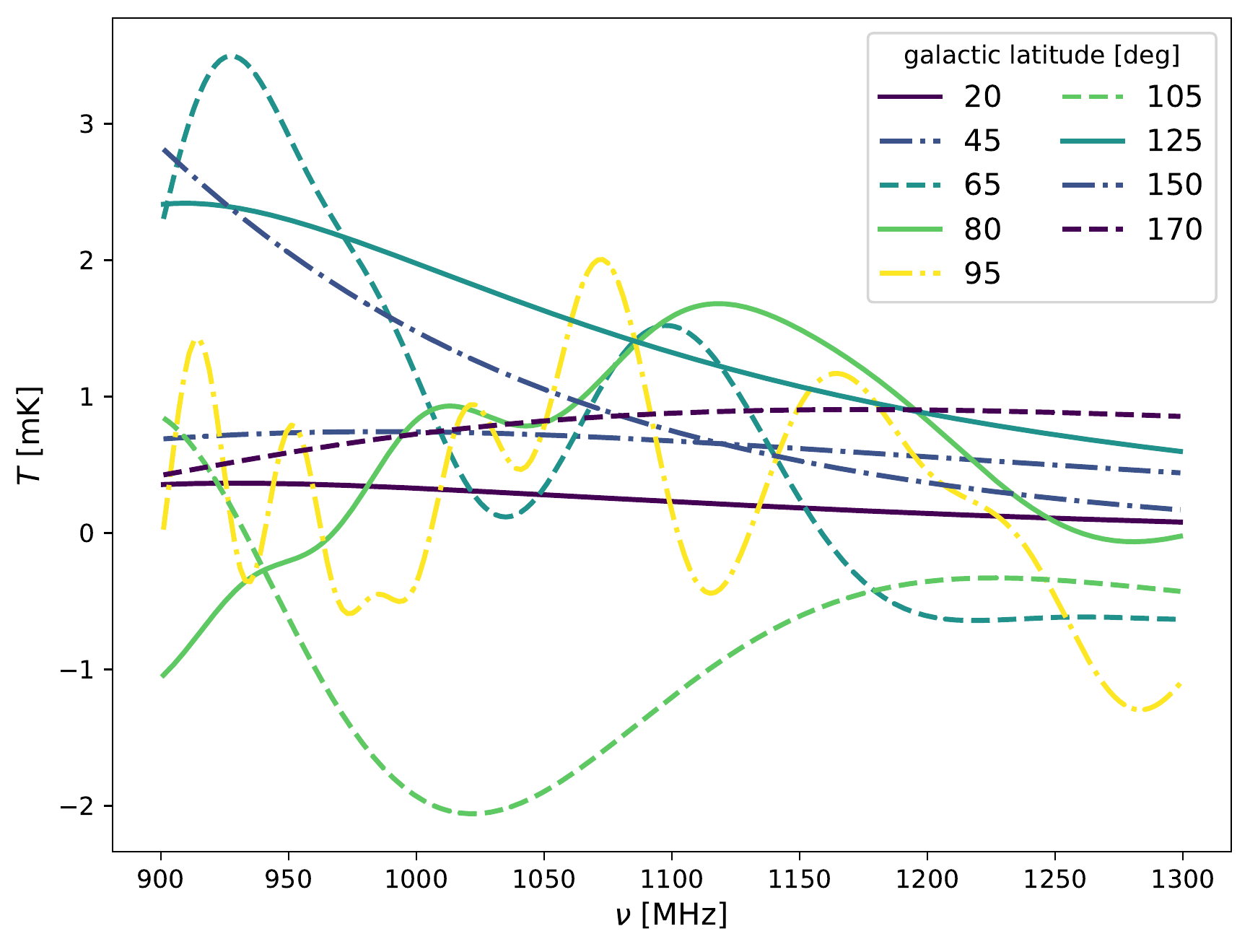}
	\caption{Temperature contribution of the polarization leakage as function of frequency, observed along different lines-of-sight at constant galactic longitude but different latitudes. The leakage has a smooth behaviour at the poles -- although dependent on line-of-sight -- and oscillates when closer to the galactic plane.}
	\label{fig:PL_LoS}
\end{figure}

While the \HI radiation is unpolarized, polarized foregrounds such as the galactic and extragalactic synchrotron emission -- additionally altered by Faraday rotation in the interstellar medium -- can spill into the unpolarized part of the received signal due to miscalibration issues \citep{Moore2013}. 

In the community, we still lack a baseline on how to model this systematic, due to lack of knowledge of the galactic synchrotron polarization at the frequencies relevant for 21\,cm intensity mapping  and of the galactic magnetic field and ionized medium where Faraday rotation happens. On-going and future surveys in polarization will help us bridge this gap \citep[e.g.][]{spass} .

Meanwhile, in the literature there are two polarization leakage models available for these frequencies, described in \citet{alonso2014_CRIME} and \citet{Shaw2015}. Even if both models are built on the same data-set (the galactic Faraday depth by \citet{Oppermann2012}), their resulting polarization leakage maps are qualitatively different (because astrophysical assumptions ought to be made even without strong supporting observational evidence). Heuristically, the \citet{alonso2014_CRIME} model outputs a more dramatic\footnote{In terms of non-smooth frequency behaviour.} leakage contamination. Therefore, we take it as a conservative guess of the true systematic and use it to complement the simulated data of this work.

Although instrument-dependent, the fraction $\epsilon$ of the spilled polarized signal is expected to be below $1\%$ \citep{santos2015}; for instance, it has been estimated to be of the order $0.6\%-0.8\%$ for the Green Bank Telescope at 800 MHz \citep{GBT_leakage}. Foreseeing  improvements in newer instruments and for updated calibration techniques, we set $\epsilon = 0.5\%$ for this study; this $\epsilon$ yields to a temperature contribution one order of magnitude higher than the cosmological signal in 21\,cm, as we will later see.

A snapshot of the polarization leakage contribution simulated with \texttt{CRIME} is in the bottom right of Fig.~\ref{fig:molls_sim} for the $\nu \in [1100 - 1102]$ channel: it has a spotty angular dependence. To appreciate its line-of-sight behaviour, we plot in Fig.~\ref{fig:PL_LoS} its equivalent brightness temperature as function of frequency, with different colours corresponding to different galactic latitudes: especially when getting closer to the galactic equator it fluctuates substantially, becoming highly pixel-dependent.

We leave for subsequent work the study of the effect of other sources of systematics, as for instance satellites contribution \citep{Harper2018_satellites} and the so-called 1/f noise \citep{Harper2018_1f}. Up to now, polarization leakage is the least explored of known systematics and has been proven to be hard to calibrate it out \citep{GBT_leakage}, in contrast with -- keeping the same examples as before --  satellite contamination that can be modelled and even avoided and the 1/f noise that can be mitigated with the scanning strategy and in the map-making process. This is why we prioritise the inclusion of the polarization leakage in the simulated data for this first GMCA study.

\subsection{Instrumental effects: telescope beam and thermal noise}

\begin{table}
	\centering
	\caption{Instrumental parameters used for computing the instrumental noise and the beam size.}
	\label{tab:instrument}
	\begin{tabular}{rcl}
		\hline
		Telescope specifics& & \\
		\hline
		dish diameter                        &$D $                     & 13.5 m        \\
		instrumental temperature     &$T_{\rm instr}$     & 20.0 K        \\
		observed fraction of the sky  &$f_{\rm sky}$        & 0.1           \\
		observation time                   &$t_{\rm obs}$        & 4000 hours  \\
		number of dishes                  &$N_{\rm dishes}$  & 64            \\
		\hline
	\end{tabular}
\end{table}

Once all the components are generated and combined, two instrumental effects are implemented to all maps: the smearing of a frequency-dependent beam and the addition of uncorrelated thermal noise.

We approximate the telescope beam with a symmetric Gaussian beam whose width depends on frequency as:
\begin{equation}\label{eq:theta}
\theta_{\rm FWHM} = \frac{c}{\nu D},
\end{equation}
with $c$ the speed of light and $D$ the telescope dish diameter. Considering the frequency range ($900 - 1300$ MHz) and the dish diameter chosen (see Table~\ref{tab:instrument}), the observed maps are smeared out to $1 - 1.4 \deg$.

Approximating the telescope beam with a spherically symmetric Gaussian smoothing is what is usually done in the 21\,cm intensity mapping foreground cleaning literature. However, it is a simplistic assumption as the the presence of side lobes in the beam profile is responsible for additional mode-mixing in the data. In this respect, the spectral complexity of the leakage contribution we include in this work can be seen as a first attempt to consider a component whose behaviour is close to what we would expect with a realistic beam too. Moreover, work is on-going for adding the effect of proper beam side lobes in the simulated data \citep{MeerKATbeam}, hence we leave the issue for a next study.

We assume the instrumental noise follows a uniform Gaussian distribution over the sky, with a frequency-dependent standard deviation of: 
\begin{equation}
\sigma_N (\nu) =  T_{\rm sys}(\nu) \sqrt{\frac{4\pi f_{\rm sky}} {\Delta \nu \, t_{\rm obs} N_{\rm dishes} \Omega_{\rm beam}}} \, 
\label{eq:sigmaN}
\end{equation}
where $T_{\rm sys}(\nu)$ is the system temperature, $f_{\rm sky}$ the observed fraction of the sky, $\Delta \nu$ the channel width,  $ t_{\rm obs}$ the total survey time, $N_{\rm dishes}$ the number of dishes. The beam solid angle is related to its width as $\Omega_{\rm beam} = 1.133 \theta_{\rm FWHM}^2$. The system temperature $T_{\rm sys}(\nu)$ is the sum of the receiver temperature and the sky temperature at a given frequency, which results in a combination of the instrument temperature $T_{\rm instr}$ and the observed frequency \citep{oneil2002}: 
\begin{equation}
T_{\rm{sys}}(\nu) = T_{\rm instr}[{\rm K}] + 66 \left(\frac{300}{\nu [{\rm MHz}]}\right)^{2.55}.
\end{equation}
The instrument and survey specifications used here are based on a MeerKLASS-like survey \citep{MeerKLASS} and summarised in Table~\ref{tab:instrument}. We generate full-sky noise maps using equation~(\ref{eq:sigmaN}) as variance per pixel.

\subsection{Observed temperature cubes}
\label{sec:datacubes}

\begin{figure}
	\includegraphics[width=\columnwidth]{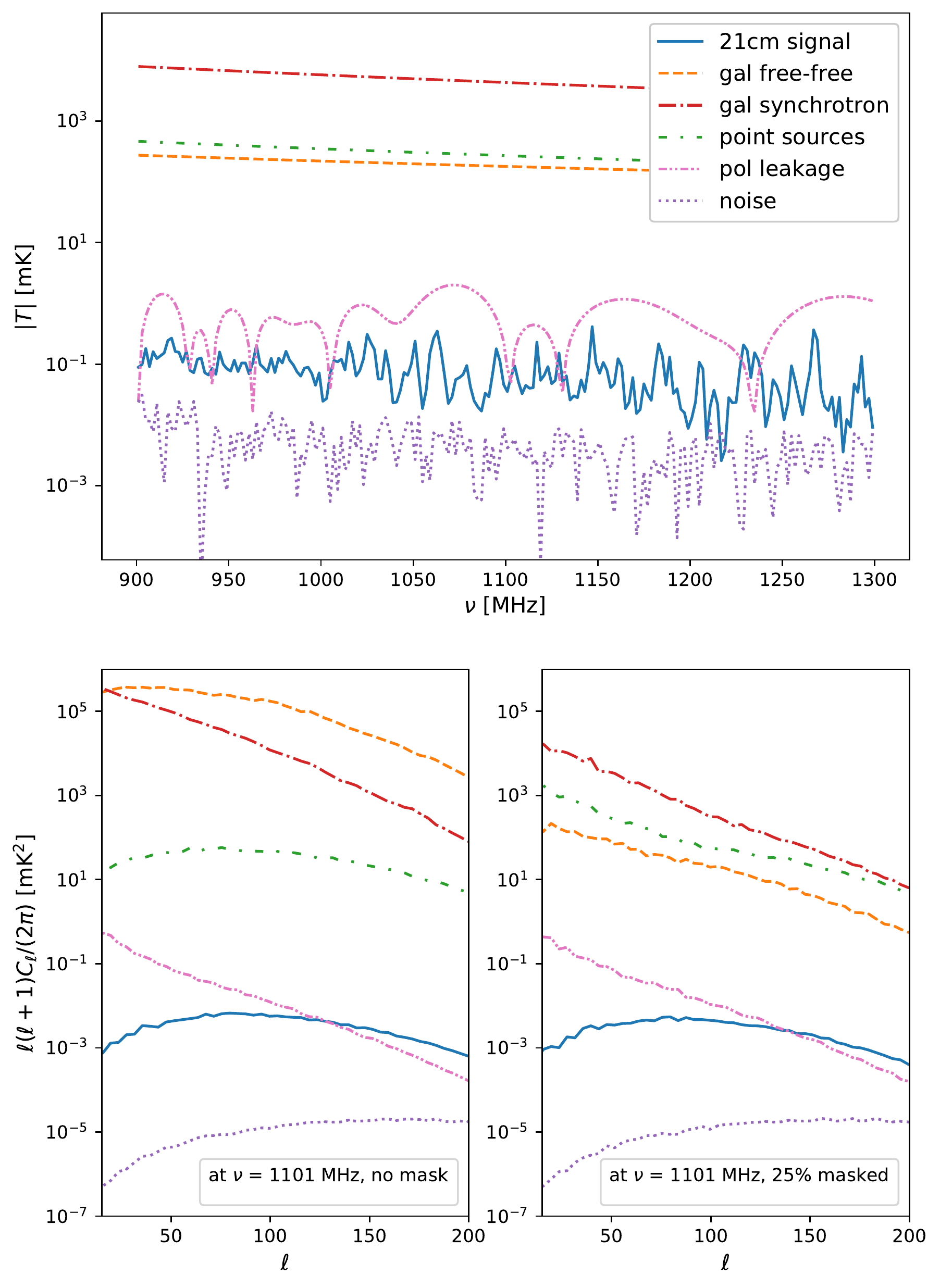}
	\caption{Components of the simulation. Top panel: brightness temperature as a function of frequency, observed along a random line of sight at galactic latitude of $95\deg$. Bottom panel: angular power spectra of channel $\nu \in [1100 - 1102]$~MHz ($z \approx 0.3$) of the full-sky maps (left) and of the $75\%$ of the sky (right) when a mask has been applied. Removing the brightest pixels at the galactic equator, where most of the galactic free-free emission is concentrated, changes the contribution of the different contaminants. The data has been smoothed as discussed in the text, suppressing power at small scale.
	}
	\label{fig:sim203}
\end{figure}

We summarise the different simulated components in Fig.~\ref{fig:sim203}: in the top panel their  temperature contributions are plotted as a function of frequency along a random line-of-sight; the polarization leakage (pink dash-dotted line) clearly sticks out among the foregrounds, as the others are indeed smooth and order of magnitudes above the cosmological signal (solid), that looks as noisy as the instrumental noise (dotted).

For each channel -- or correspondingly frequency or redshift --  we sum the maps of the 21\,cm  cosmological signal \textbf{C}  with those of all foregrounds \textbf{F}, we convolve the total temperature map with the frequency-dependent beam, we add the white noise \textbf{N}. This makes up the observed data-cube. According to the mixture model in Equation \ref{eq:master}, the temperature maps are all assumed to be at the same resolution. For that purpose, we further re-convolve all maps appropriately to let them all share the same resolution, i.e. that of the lowest frequency channel where the beam $\theta_{\rm FWHM}$ is the largest. Schematically:
\begin{equation} \label{eq:final_maps}
\mathbf{X} = [ (\mathbf{C}  + \mathbf{F} )  * B + \mathbf{N}]* (B_{\rm low} - B)\,.
\end{equation}

Having all temperature maps at the same resolution is not essential. We also perform the source separation without the additional deconvolution, but we typically have to set a higher number of sources $n_{\rm s}$ for reaching a satisfactory foreground cleaning -- compared to the case where all maps share the same resolution -- thus risking to over-clean and miss true signal \textbf{C} in the residuals. This is expected as the mixture model of the signal becomes more complex due to the frequency-dependent effect of the beam. So we take into account the latter with the deconvolution $(B_{\rm low} - B)$.

Ultimately, this further deconvolution is a loss  of smallest angular information available in the observed maps. We do not discuss this issue here, since a version of GMCA that performs the beam deconvolution at the same time as the blind source separation has been tested on two-dimensional data \citep[decGMCA,][]{decGMCA} and effort is ongoing for extending decGMCA on data sampled on the sphere \citep{remi_decGMCA}; thus, the results of this paper would generically hold for a decGMCA application, with the advantage of retaining the fully-available small scale information. 

The simulation spans a frequency range of $\nu \in  [900 - 1300]$ MHz, corresponding to redshift $z \in [0.09 -0.58]$. We slice the data-cubes in  bins of $\Delta \nu =2,\,5\,{\rm and}\,10$ MHz, corresponding to numbers of channels $n_{\rm ch} = 200,\,80\,{\rm and}\,40$: in this way we can test the dependence of  GMCA performance on $\Delta \nu $ and on $n_{\rm ch}$.

\section{The pipeline}
\label{sec:pipeline}

\subsection{Recovering the input signal}

\begin{figure}
	\includegraphics[width=\columnwidth]{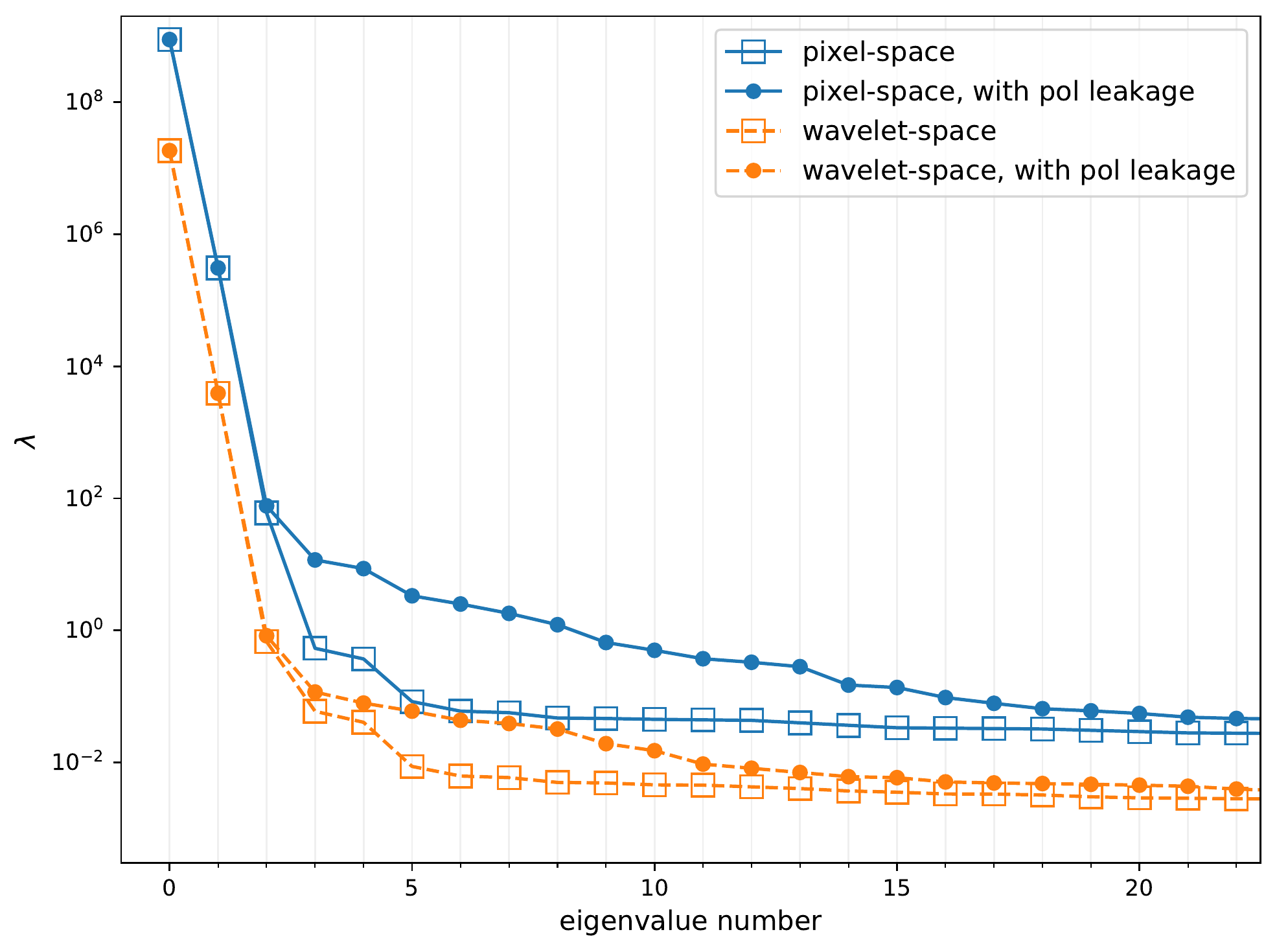}
	\caption{Principal eigenvalues of the frequency covariance matrix of the $n_{\rm ch} = 200$ simulation, using the standard pixel-space data-cube \textbf{X}  (blue lines) and  its wavelet-transformed counterpart \textbf{X}$^{\rm wt}$ (orange). Empty squares (filled circles) correspond to the scenario without (including) the polarization leakage component. The contribution of a component like the polarization leakage spreads the foregrounds through a larger number of eigenvalues, making foreground cleaning more problematic, i.e. a larger number of  degrees of freedom should be eliminated. On the other hand, working in wavelet-space restricts the spreading in fewer degrees of freedom and makes the contamination more tractable.
	}
	\label{fig:eigenv}
\end{figure}

As anticipated in Section \ref{sec:GMCA}, GMCA looks for components of the signal that are sparse in the wavelet domain. Thus, once the data-cube \textbf{X} of equation~(\ref{eq:final_maps}) has been assembled and its mean removed channel-wise, we wavelet-transform it, obtaining \textbf{X}$^{\rm wt}$. 
By looking at the principal eigenvalues of the covariance matrix of the data-cubes \textbf{X} and \textbf{X}$^{\rm wt}$ ( Fig.~\ref{fig:eigenv}, symbols in blue and orange respectively), we can already appreciate the advantage of running the blind source separation in wavelet-space rather than pixel-space: the transition between highly correlated modes in frequency (foregrounds) and uncorrelated modes (signal and noise) happens for smaller eigenvalue number for the wavelet case; the latter is especially true when we add a component like polarization leakage (dots versus squares in Fig.~\ref{fig:eigenv}), that mixes the modes of the covariance matrix and makes the transition among them smoother.

 GMCA promotes sparsity in the decomposition process of \textbf{X}$^{\rm wt}$, as in equation~(\ref{eq:gmca}), and estimates the mixing matrix $\mathbf{\tilde{A}}$. We determine the foreground components that GMCA identifies, \textbf{X} $^{\rm GMCA}$, by projecting the input data \textbf{X} on  $\mathbf{\tilde{A}}$; the cleaned maps \textbf{X}$^{\rm cleaned}$ are the residuals of the GMCA source separation:
\begin{equation} \label{eq:Xclean}
\mathbf{X}^{\rm cleaned} = \mathbf{X} - \mathbf{X}^{\rm GMCA} = \mathbf{X}- \mathbf{\tilde{A}} (\mathbf{\tilde{A}}^T \mathbf{\tilde{A}})^{-1} \mathbf{\tilde{A}}^T \mathbf{X}\,.
\end{equation}

\subsection{GMCA performance}

\textbf{X}$^{\rm cleaned}$ of equation~(\ref{eq:Xclean}) are the maps that would be analysed for extracting science in a real context. In next paragraphs, we show how we evaluate the performance of the foreground cleaning by comparing \textbf{X}$^{\rm cleaned}$ with the input data, i.e. the cosmological signal {\it and} the instrumental noise  \textbf{C}$+$\textbf{N}.

\subsubsection{Power spectrum estimation}
\label{sec:pk}

The \HI intensity two-point statistics carries a great deal of the cosmological information, as for any tracer of the underlying matter field. Hence, the performance of a given foreground cleaning method should be assessed at least in terms of its ability to recover the true power spectrum at different radial and angular scales.

\paragraph{Angular scales.}

Since the observed temperature data \textbf{X} is sampled on spheres of $n_{\rm pix}$ pixels, for a  shell at fixed frequency $\nu$, it is convenient to expand its distribution $\Delta T (\nu) = \textbf{X}_{\nu} - \langle  \textbf{X} \rangle_{\nu}$  in spherical harmonic functions $Y_{\ell m}(p)$. We estimate the harmonic coefficients as a summation over the pixels $p$ of the map:
\begin{equation}
a_{\ell m} (\nu) = \sum_{p = 1}^{n_{\rm pix}} \Delta T (\nu, p) Y^*_{\ell m}(p)\,.
\end{equation}
All the above holds for any temperature data-cube we assess, i.e. we can substitute $\textbf{X}$ with foregrounds \textbf{F} or 21\,cm cosmological signal \textbf{C} and so on.
The angular power spectrum is defined as $C_\ell \equiv \langle |a_{\ell m}|^2 \rangle$. For calculating the $C_\ell$ of each map, we make use of the software package \texttt{NaMaster}\footnote{\url{https://github.com/LSSTDESC/NaMaster}}, whose algorithm is described in \citet{NaMaster}. \texttt{NaMaster} is a pseudo-$C_{\ell}$ estimator that can also efficiently take into account incomplete sky coverage, as it will be the case in  Section~\ref{sec:mask}.

For instance, in the bottom panels of Fig.~\ref{fig:sim203} we plot the angular power spectra of all components of the simulation sampled at frequency $\nu = 1101$ MHz; in the left, the $C_{\ell}$ have been computed for the full-sky, in the right we have first applied a mask covering the equatorial $25\%$ of map. The power amplitude of the astrophysical foregrounds is up to $7-8$ orders of magnitude higher than the 21\,cm signal in both panels; what change in the two cases are the individual foregrounds components spectra and their hierarchy, as consequence of the different spatial features of the foregrounds --  morphological differences that will help GMCA to detect them. On the contrary, the $C_{\ell}$ of the cosmological signal and noise do not change among panels, showing that they are spatially isotropic.

For each channel map, we compare the angular power spectrum of \textbf{X}$^{\rm cleaned}$, that we dub $C_{\ell}^{\rm cleaned}$, with that of the input $\textbf{C} + \textbf{N}$, i.e. $C_{\ell}^{\rm true}$. Averaging over all channels of the simulation, we build the quantity: 
\begin{equation} \label{eq:Rl}
R_{\ell} = \Big  \langle \frac{C_{\ell}^{\rm true} - C_{\ell}^{\rm cleaned}}{C_{\ell}^{\rm true}}  \Big \rangle_{\rm chs}\,.
\end{equation}
We will use $R_{\ell}$ to assess the performance of GMCA in different scenarios.

\paragraph{Radial scales.}

\begin{figure}
	\includegraphics[width=\columnwidth]{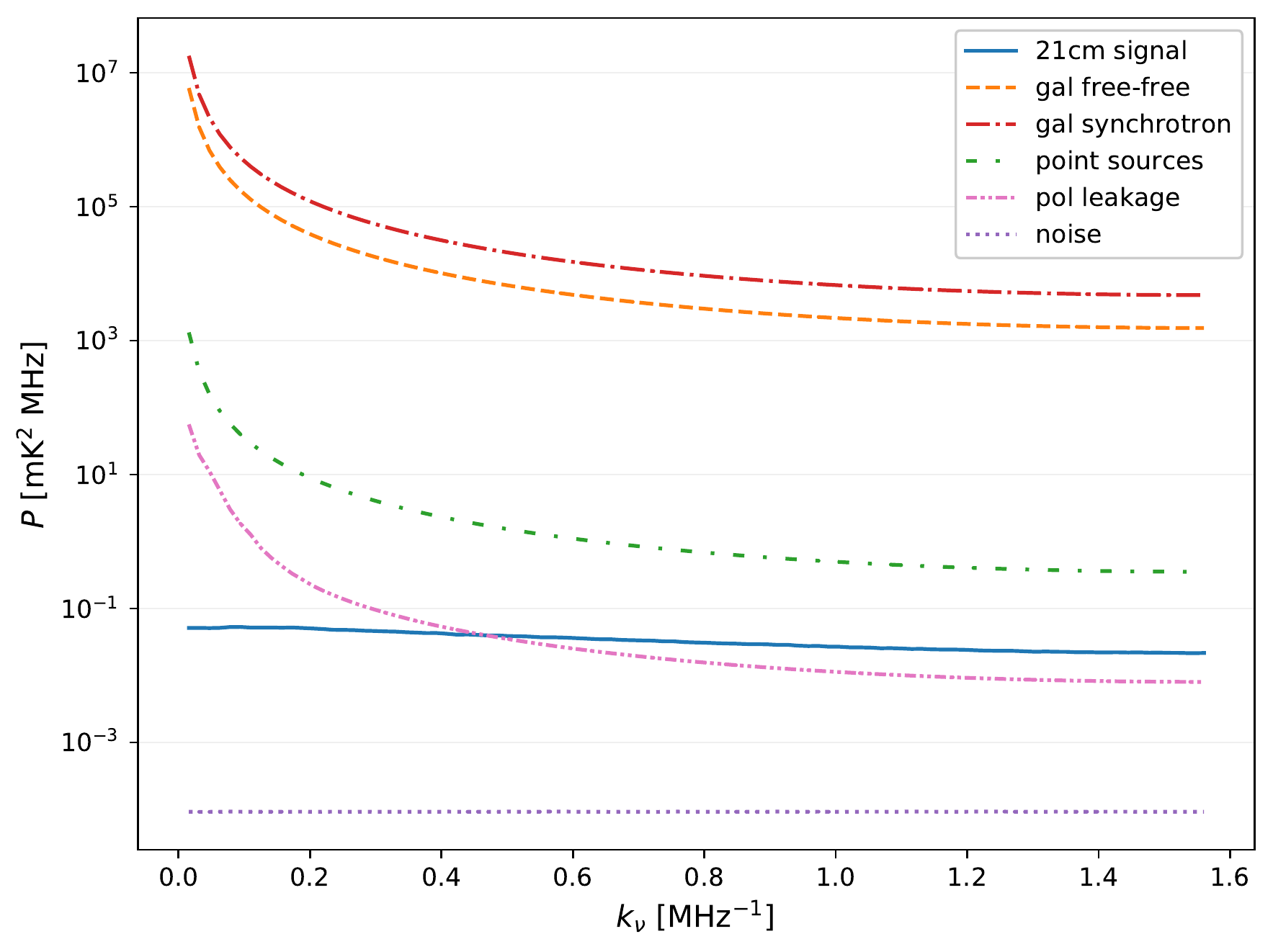}
	\caption{Radial power spectra in frequency-space for all components of the $n_{\rm ch} = 200$ simulation, being smoothed by the telescope beam.
	}
	\label{fig:pknu_sim}
\end{figure}

Given the spectral nature of the 21\,cm signal, the possibility of achieving unprecedented redshift resolution while sampling big volumes is one of the characteristics that makes intensity mapping highly appealing for cosmology. In this sense, it is crucial to investigate that also the radial direction information is retrieved after the cleaning process.

To estimate the two-point statistic in the radial direction, one could either rely on the angular cross-correlation of maps at different redshifts  to avoid dealing with light-cone effects and curved-sky issues \citep{Montanari2012,Asorey2012}, or otherwise proceed with defining a one-dimensional $k_{\parallel}$ power spectrum estimator  \citep{alonso2014_CRIME,Villaescusa2017,blake2019}.
Here we opt for a simpler approach: we compute the one-dimensional power spectrum directly in frequency space, $P(k_{\nu})$ with  $k_\nu = \frac{2\pi}{\nu}$. This choice makes difficult a direct comparison with cosmological observables, nevertheless it supplies a straightforward insight about the efficiency and deficiencies of foreground cleaning in the radial direction. In practice:
\begin{enumerate}
	\item for each line-of-sight (i.e. pixel), we Fourier-transform the $\Delta T(\nu)$ field along $\nu$
		\begin{equation}
			\tilde{\Delta T}(k_\nu) = \int d\nu \,\Delta T(\nu)\, e^{-ik_\nu \nu}  \,;
		\end{equation}
	\item we compute the power spectrum
			\begin{equation}
					P(k_\nu) = \Delta\nu\, \langle |\tilde{\Delta T}(k_\nu,p)|^2 \rangle \,,
			\end{equation}
	by averaging over all the lines-of-sight $p$.
\end{enumerate}

In Fig.~\ref{fig:pknu_sim} we show the  $P(k_\nu)$ for each component of the simulation. As for the $C_\ell$, the amplitude of the foregrounds $P(k_\nu)$ is by far higher than that of the cosmological signal, the one of the noise is negligible. The high correlation in frequency of the foregrounds is also evident in their $P(k_\nu)$, that sharply increase towards higher frequency scales, in contrast with the 21\,cm signal that -- as the noise -- displays a flat  $P(k_\nu)$, with a slow decrease for high $k_\nu$ due to the effect of the beam smoothing \citep{Villaescusa2017}.

In the same fashion of equation~(\ref{eq:Rl}), we define the quantity $R_{\nu}$ for comparing the input and reconstructed radial power spectra:
\begin{equation} 
R_{\nu} = \frac{P(k_\nu)^{\rm true} - P(k_\nu)^{\rm cleaned}}{P(k_\nu)^{\rm true}} \,.
\end{equation}

\subsubsection{Residual projection}

Two contributions make the cleaned maps  go astray from the input $\textbf{C} + \textbf{N}$: (i) foregrounds are not fully captured in \textbf{X}$^{\rm GMCA}$, contaminating \textbf{X}$^{\rm cleaned}$, and (ii) true cosmological signal partly leaks into \textbf{X}$^{\rm GMCA}$ and is lost. To quantify those effects individually, we define the residual projections.

The foreground residual that leaks into the recovered signal and noise is:
\begin{equation} \label{eq:F_res}
\textbf{X}_R^F = \textbf{F} - \mathbf{\tilde{A}} (\mathbf{\tilde{A}}^T \mathbf{\tilde{A}})^{-1} \mathbf{\tilde{A}}^T \textbf{F}\,,
\end{equation}
where \textbf{F} is the input foregrounds data-cube, from which we subtract the foreground maps projected on to the GMCA-estimated mixing matrix $\mathbf{\tilde{A}}$. Similar to equation~(\ref{eq:F_res}), we define the signal plus noise, $\textbf{C} + \textbf{N}$, that leaks in the estimated foregrounds as:
\begin{equation} \label{eq:CS_lkg}
\textbf{X}_F^{CN} = \mathbf{\tilde{A}}(\mathbf{\tilde{A}}^T \mathbf{\tilde{A}})^{-1} \mathbf{\tilde{A}}^T (\textbf{C}+\textbf{N})\,.
\end{equation}

The foreground removal succeeds when the power spectra of both \textbf{X}$_R^F$ and \textbf{X}$_F^{CN}$ are negligible compared to that of $\textbf{C} + \textbf{N}$.

\section{Results}
\label{sec:results}

\begin{figure*}
	\includegraphics[width=\textwidth]{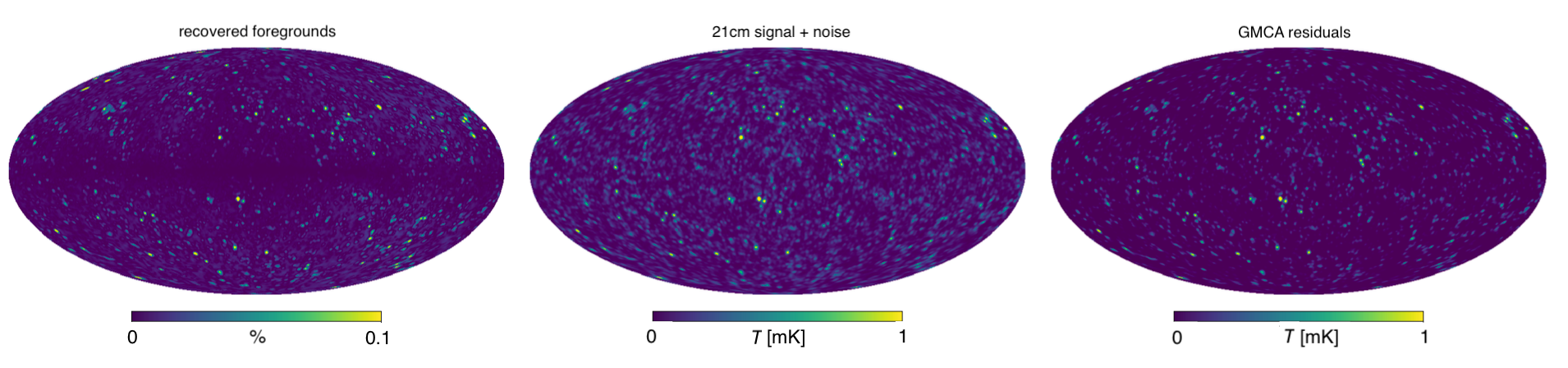}
	\caption{Foreground removal  for the  $\nu \in [1100 - 1102]$~MHz channel, GMCA has looked for $n_{\rm s} = 3$ components. Left map: relative difference in percentage between the input foregrounds, \textbf{F},  and what found by GMCA,  \textbf{X}$^{\rm GMCA}$. Middle map: input 21\,cm signal and instrumental noise, $\textbf{C} + \textbf{N}$. Right map: foreground removal residuals \textbf{X}$^{\rm cleaned}$, to be compared with middle map. Foregrounds total temperature is recovered at sub-percent level, especially in the galactic plane, yet we miss some of the low temperature features of the input signal map; the brightest 21\,cm spots are recovered in the map in the right, and they also represent the regions at most off-set in the foreground recovery (left).
	}
	\label{fig:residuals_molls}
\end{figure*}

To better understand the foreground removal problem, we first run GMCA on data-cubes with foreground contributions of galactic synchrotron and free-free diffuse emissions and extragalactic diffuse emission and point sources; we later increase the degree of complexity of the foregrounds by adding the polarization leakage. This first assessment makes also possible a more direct comparison of GMCA with other methods in the literature. 

We study how the number and thickness of the channels affect the performance of the foreground cleaning; we further assess its performance when some channels are missing altogether, which is often the case in real surveys; we check whether masking the pixels with higher foreground contamination eases the cleaning task; we eventually add the polarization leakage in the game and, lastly, we try the same tasks with another source separation algorithm -- FastICA -- for comparison.

We start by visually inspecting the GMCA-reconstructed maps. We feed the 200 channel data-cube (missing the polarization leakage contribution) to GMCA setting to $n_{\rm s} =3$ the number of morphologically different sources to search. In Fig.~\ref{fig:residuals_molls} are the results for the $\bar{\nu} = 1101$ MHz channel: we show the sky mollweide projections of {(i) \it left panel:} the difference in intensity between the input foregrounds \textbf{F} and what is identified by GMCA, i.e. $(1-\textbf{X}^{\rm GMCA}/ \textbf{F})_{\bar{\nu}}$; {(ii) \it middle panel:} the input signal and instrumental noise, $(\textbf{C}+\textbf{N})_{\bar{\nu}} $; {(iii) \it right panel:}   the cleaned map recovered with GMCA, \textbf{X}$^{\rm cleaned}_{\bar{\nu}}$. Looking at the left panel: GMCA has remarkably identified the true intensity of the foregrounds with sub-percent level of accuracy. Is this achievement enough for the recovery of the feeble 21\,cm signal? We compare the remaining panels: the input $\textbf{C} + \textbf{N}$ (middle) with the output \textbf{X}$^{\rm cleaned}$ (right). The bright spots where emission is greatest are clearly present in both maps, however much of the fainter features present at all scales in the $\textbf{C} + \textbf{N}$ map are missing in \textbf{X}$^{\rm cleaned}$. Next we will plot the power spectra of these maps to assess the information that we can still safely extract from \textbf{X}$^{\rm cleaned}$. Inspecting further the maps of Fig.~\ref{fig:residuals_molls}, we notice that the pixels where foregrounds are worse caught correspond to the bright spots of the true signal outside the galactic plane, whereas the galactic plane pixels, where foregrounds more strongly shine, correspond to  those pixels that -- counter-intuitively --  experience the best recovery of the foreground emission. The latter remark will be further corroborated in Section~\ref{sec:mask} where we perform the foregrounds cleaning after applying masks to the maps.

\subsection{The dependence on the number of channels}

\begin{figure}
	\includegraphics[width=\columnwidth]{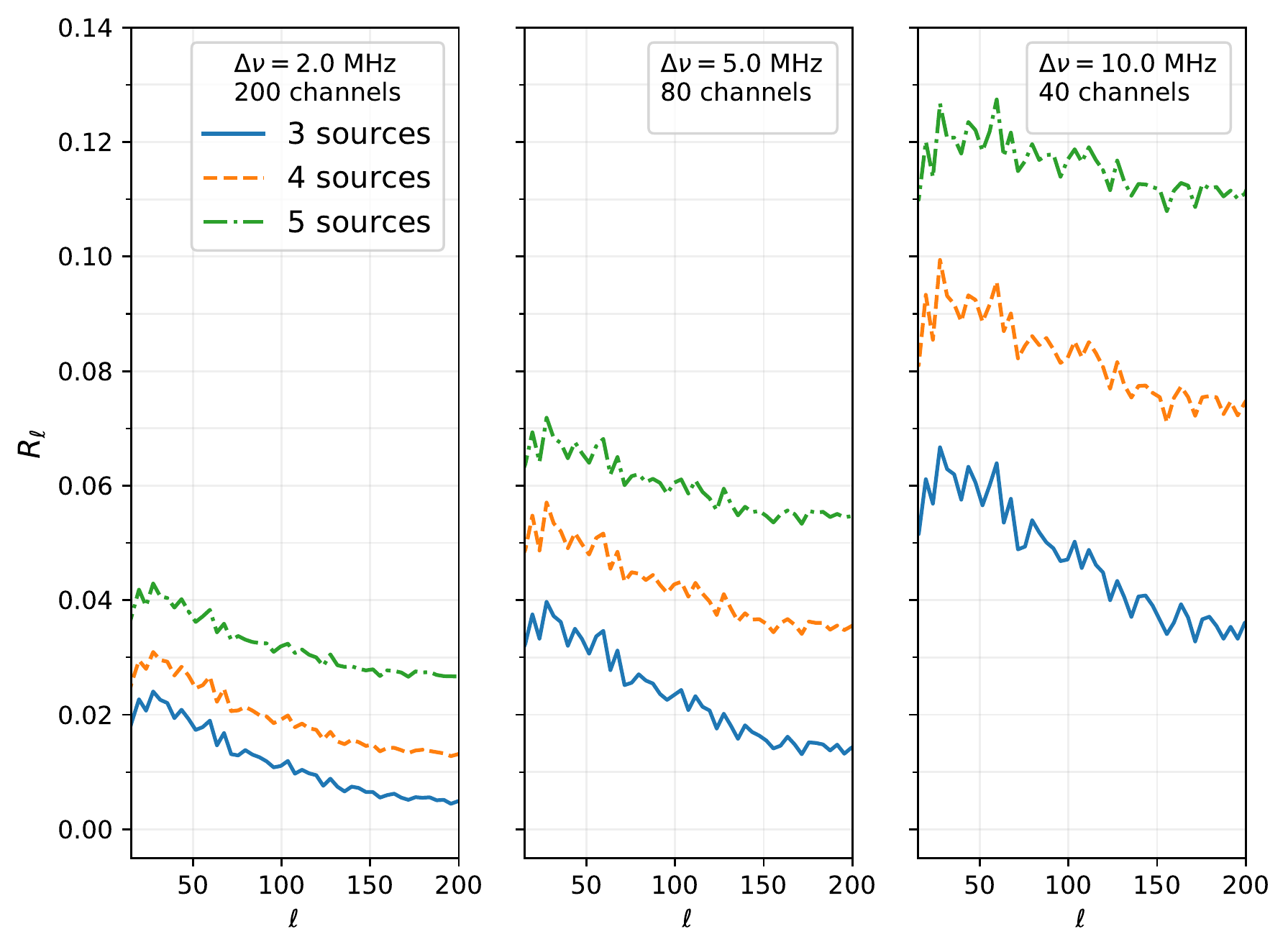}
	\caption{Relative difference in angular power spectrum between the GMCA recovered signal and the input cosmological signal and instrumental noise, averaged among all channels. The 3 different simulations (panels) are full-sky and GMCA has been run with number of sources $n_{\rm s} =$ 3, 4 and 5 (solid, dashed and dash-dotted lines respectively). These simulations lack the polarization leakage contribution. Overall, GMCA recovers the signal within a few percent bias in angular power spectrum.}
	\label{fig:Rl_m00}
\end{figure}

\begin{figure}
	\includegraphics[width=\columnwidth]{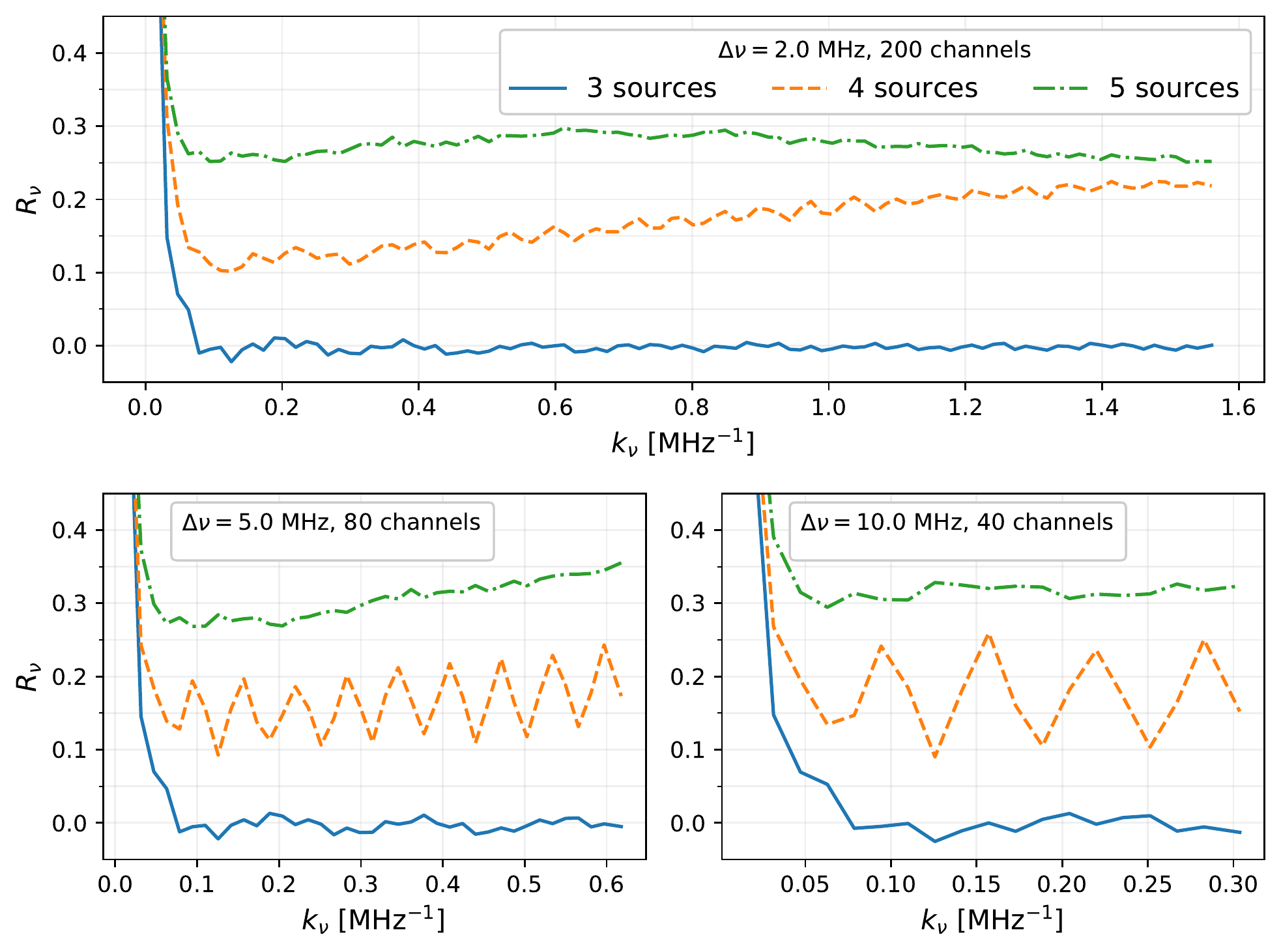}
	\caption{Relative difference in radial power spectrum between the GMCA recovered signal and the input cosmological signal and instrumental noise. The 3 different simulations (panels) are full-sky and GMCA has been run with number of sources $n_{\rm s} =$3, 4 and 5 (solid, dashed and dash-dotted lines respectively). These simulations lack the polarization leakage contribution. GMCA recovers unbiased information in the radial direction for $k_\nu \gtrsim 0.05$ MHz$^{-1}$ when $n_{\rm s} =3$ is set.
	}
	\label{fig:Rnu_00}
\end{figure}

\begin{figure}
	\includegraphics[width=\columnwidth]{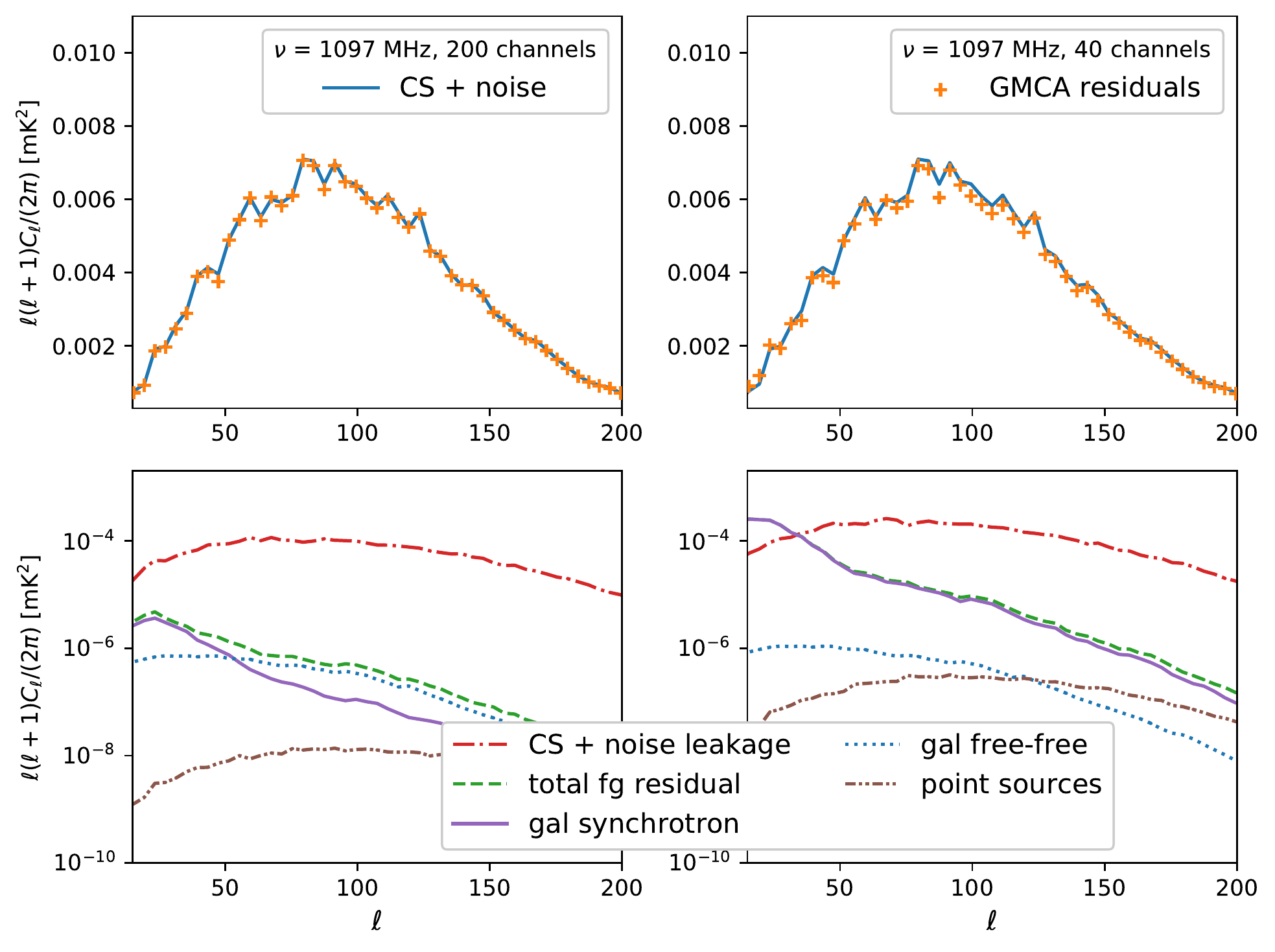}
	\caption{GMCA results for the  $\nu \in [1096 - 1098]$~MHz channel of the full-sky simulation with no polarization leakage, using $n_{\rm s} = 3$ number of components. In the top panels are the angular power spectra of the input cosmological signal and noise $\textbf{C} + \textbf{N}$ (solid line) and the GMCA recovered signal \textbf{X}$^{\rm cleaned}$ (plus signs). In the bottom panels are the leakage of the input cosmological signal and noise into the GMCA found foregrounds \textbf{X}$_F^{CN}$ (dash-dotted line), and the residuals of the input foregrounds left over in the GMCA recovered signal \textbf{X}$_R^F$  (dashed line) and for the individual foreground component (see legend). The left panel corresponds to a GMCA run on the full 200 channels available, in the right using only 40 consecutive channels of the simulation. With less channels, i.e. less frequency information to characterise the foregrounds, GMCA performs worse, especially for identifying the galactic synchrotron (featuring a spatially-varying spectral index), whose residual leaks in the recovered signal at large scales.
	}
	\label{fig:40channels_res}
\end{figure}

For each simulation set-up previously described (with different number and thickness of channels), we perform various foreground removals with GMCA varying the number of sources $n_{\rm s}$. We compute angular and radial power spectra of all cleaned maps, and compare with the ground truth ones, as described in Section~\ref{sec:pk}. We show the angular power spectra relative difference $R_{\ell}$ in Fig.~\ref{fig:Rl_m00} and the radial counterpart $R_{\nu}$ in Fig.~\ref{fig:Rnu_00}. We plot these quantities for the simulation with $n_{\rm ch} = 200$ and $\Delta\nu = 2$ MHz, $n_{\rm ch} = 80$ and  $\Delta\nu = 5$ MHz and $n_{\rm ch} = 40$ and $\Delta\nu = 10$ MHz  in panels from left to right (from top to bottom) in Fig.~\ref{fig:Rl_m00} (Fig.~\ref{fig:Rnu_00}). Different lines correspond to $n_{\rm s} =$ 3, 4 and 5 (solid, dashed, dot-dashed respectively).

Focusing on Fig.~\ref{fig:Rl_m00}, in the best scenario -- 200 channels and $n_{\rm s} = 3$ -- the angular power spectrum of $\textbf{C} + \textbf{N}$ is recovered on average with a $2\%$ bias on large scales down to  $0.5\%$ for $\ell>150$. Setting $n_{\rm s}$ to higher values leads $R_{\ell}$ to increase in amplitude, for instance doubling for the $n_{\rm s} = 5$ case. Having a lower number of channels also impacts negatively $R_{\ell}$, as its amplitudes in the middle and right panels are higher and go up to $12\%$ for the $n_{\rm ch} = 40,\,n_{\rm s} = 5$ case. We expect this as, even if the simulations cover the same range in frequency, the higher number of channels/maps, the larger the data-set GMCA can rely on for extracting components and mixing matrix.

Also assessing the information in the radial direction, results are promising: setting $n_{\rm s} = 3$ we recover the power spectrum in frequency space within few percent (solid lines in all panels of Fig.~\ref{fig:Rnu_00}). This bias increases up to $\approx 35\%$ when increasing $n_{\rm s}$, showing some mild scale-dependence. In contrast to the angular $R_\ell$, the results in $R_{\nu}$ happen to be quite $n_{\rm ch}$-independent; of course the smallest scale we can reach in $k_\nu$ is dictated by the frequency-resolution $\Delta \nu$ of the simulations (the highest wave-number to be trusted is $\pi \,n_{\rm ch} / \Delta \nu$), nonetheless $R_\nu$ is under control for $k_\nu \gtrsim 0.05$ MHz$^{-1}$ for all $\Delta \nu$ scenarios. Ignoring light-cone effects, we can crudely relate $k_\nu$ to $k_\parallel$ -- its comoving-distance counter-part: $k_\parallel \approx \frac{\nu_{\rm 21cm}\, H(z)}{c \,(1+z^2)}\, k_\nu$, with $H(z)$ the Hubble parameter; by using the cosmological parameters of the simulation and the middle redshift of the data-cube, we can claim to recover the true radial power spectrum for $k_\parallel \gtrsim 0.02\,h$ Mpc$^{-1}$. A noticeable feature of the results in Fig.~\ref{fig:Rnu_00} is the oscillating behaviour of some of the $R_\nu$ displayed: it is due to {\it ringing} effects in computing the Fourier-transforms because of the presence of numerical zeros in the $\Delta T$ data, originated when subtracting the map mean from pixels whose values were close to the mean; we explicitly checked that those effects disappear when we apply an additional and more aggressive smoothing on the liable maps, converging to a still $R_\nu$.

Looking at  Fig.~\ref{fig:Rl_m00}, we confirm the expectation that the larger the number of channels available, i.e. the more the data, the better the GMCA performance at characterising the foregrounds. However, since the three different simulations cover the same frequency range, a different number of channels leads to a different thickness $\Delta \nu$ of channels: could this latter parameter play a role in the way GMCA works? The angular power spectra of $\textbf{C} + \textbf{N}$ is higher for thinner channels, because of the higher instrumental noise but manly because of purely geometrical considerations (e.g. the $C_\ell$ of $\textbf{C} + \textbf{N}$ for the $\Delta \nu = 2$ MHz case is roughly 40 times higher than in the $\Delta \nu = 10$ MHz case).

To clarify the role of both $n_{\rm ch}$ and $\Delta \nu$ in the foreground cleaning, we perform the following exercise. We run GMCA using only a sample of $40$ consecutive channels of both the $n_{\rm ch} = 200$ and $n_{\rm ch} = 80$ channel simulations: the level of $R_\ell$ increases by 5 and 3 times respectively and independently of $n_{\rm s}$, compared to the results in Fig.~\ref{fig:Rl_m00}. It is thus clear that GMCA struggles more when it has access to less channels, independently of $\Delta \nu$. Moreover, remarking that (i) with 40 consecutive channels of the $\Delta \nu = 2$ MHz simulation the situation worsens more than with 40 consecutive channels of the $\Delta \nu = 5$ MHz simulation, and (ii) in both cases the performance of GMCA is worse than with the full 40 channel simulation with $\Delta \nu = 10$ MHz (right panel of Fig.~\ref{fig:Rl_m00}), points to the importance of the span in frequency of the data-cube for a successful foreground removal. We will come again to the same conclusion when we will try GMCA on cropped data in the RFI Section~\ref{sec:RFI}: regardless of $\Delta \nu$, it is better to have GMCA working on the full frequency range available even when channels are missing. Instead, we find no strong arguments for aiming to a specific channel width, as far as it concerns the GMCA reconstruction.

To show how compelling are the span in frequency of the data-cube and the number of channels we work with, we plot in Fig.~\ref{fig:40channels_res} the results of the same $\Delta\nu = 2$ MHz channel (middle frequency $\nu = 1097$ MHz) when GMCA has run on the whole 200 channel data (left column), or just on a 40 channel subset (right column). All curves are angular power spectra: solid blue is the input $\textbf{C} + \textbf{N}$ and with orange plus signs we plot \textbf{X}$^{\rm cleaned}$; other colours and line styles refer to the projections of the leaked signal \textbf{X}$_F^{CN}$, of the total residual foregrounds \textbf{X}$_R^F$ and of residuals of single foreground components. The change in amplitude of the projection of the residual galactic synchrotron (solid violet lines in the lower panels) is evidence that, for the very same channel, GMCA characterises synchrotron more poorly in the case on the right with the only difference being the smaller number of channels used and frequency span covered. 

We choose $n_{\rm ch} = 200$ to be the reference simulation in the rest of the analysis.

\subsection{Selecting $n_{\rm s}$}
\label{sec:convergence}

\begin{figure*}
	\includegraphics[width=\textwidth]{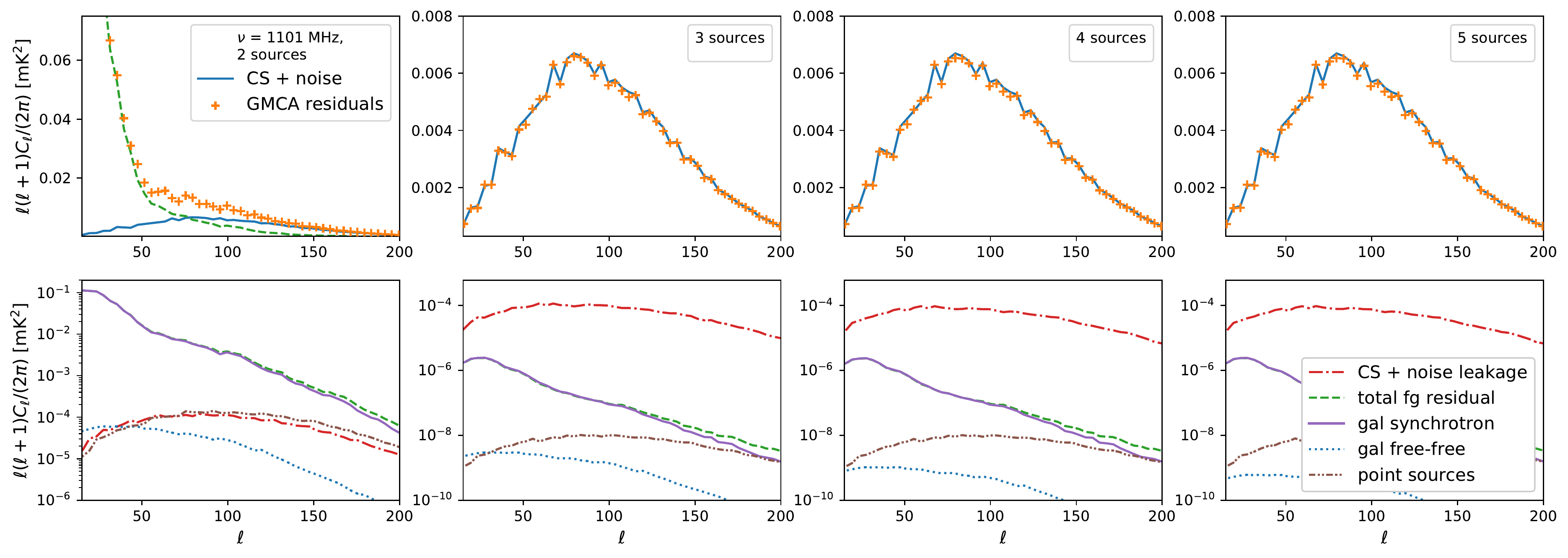}
	\caption{GMCA results for the  $\nu \in [1100 - 1102]$~MHz channel of the  full-sky simulation with no polarization leakage, using $n_{\rm s} =$ 2, 3, 4 and 5 number of components (panels from left to right). In the top panels are the angular power spectra of the input cosmological signal and noise $\textbf{C} + \textbf{N}$ (solid line) and the GMCA recovered signal \textbf{X}$^{\rm cleaned}$ (plus signs). In the bottom panels are the leakage of the input cosmological signal and noise into the GMCA found foregrounds \textbf{X}$_F^{CN}$ (dash-dotted line), and the residuals of the input foregrounds left over in the GMCA recovered signal \textbf{X}$_R^F$ (dashed line) and for the individual foreground components. There is a clear convergence in the foreground removal setting $n_{\rm s} = 3$ and higher.
	}
	\label{fig:sim203_ich100}
\end{figure*}

From Figs.~\ref{fig:Rl_m00} and ~\ref{fig:Rnu_00} it is clear that setting the sources GMCA looks for to $n_{\rm s}=3$ is optimal for the foregrounds contribution, sky-coverage and frequency-range set-ups we are considering, leading to an unbiased recovery of the information in the radial direction and within few percent in the perpendicular one. However, we will contradict this result later in the analysis, when masking the brightest pixels of the maps or adding a mode-mixing component as the polarization leakage. Practically, we cannot expect a specific values of $n_{\rm s}$ to hold in general because of the variety of realistic survey scenarios, which would be impossible to simulate perfectly, also taking into account the addition of unknown systematics or astrophysical contributions that could actually manifest in the observations and our ignorance of the 21\,cm signal itself; moreover, specifically concerning how GMCA works, we cannot rely on the same level of sparsity when considering different regions of the sky and maps with different resolutions.

Indeed, when dealing with real data, it has been removed order $\sim 10$ or more number of sources/ independent components/ principal modes  \citep{Chang2010,Masui2013,Switzer2013,wolz2017,anderson2018}.

How to set the number of sources the blind source separation algorithm has to look for, having no ground truth to compare against? A good starting point is to look at the eigenvalues of the covariance matrix of the signal in the domain we work in, as we do in Fig.~\ref{fig:eigenv}, although, especially with the inclusion of mode-mixing components (filled dots), it can be problematic to distinguish foregrounds modes from cosmological ones.

Here, we have a closer look at results of the simplified scenario (no polarization leakage) to check whether we could tell a priori $n_{\rm s}=3$ is optimal by looking at the recovered power spectra. We will later check if we will be able to apply what we learn in this simplified scenario in more complex ones.

We look more closely at how the GMCA performance changes when we vary $n_{\rm s}$. In Fig.~\ref{fig:sim203_ich100} we plot GMCA results for just one channel (of central frequency $\nu = 1101$ MHz); columns refer to different runs of GMCA where the number of sources has been set to $n_{\rm s}=$ 2, 3, 4 and 5 from left to right. All curves are angular power spectra: solid blue is the input $\textbf{C} + \textbf{N}$ and with orange plus signs we plot \textbf{X}$^{\rm cleaned}$; other colours and line styles refer to the projections of the leaked signal \textbf{X}$_F^{CN}$ (red dash-dotted), of the total residual foregrounds \textbf{X}$_R^F$ (green dashed) and of residuals of single foreground components (these plots have same structure and colour-coding of Fig.~\ref{fig:40channels_res}). The behaviour of the \textbf{X}$^{\rm cleaned}$ spectrum changes abruptly from the $n_{\rm s}=$ 2 to 3 case (first two columns from left), whereas it stays stationary for the remaining $n_{\rm s}=$ 4 and 5 cases. In the $n_{\rm s}=2$ run, \textbf{X}$^{\rm cleaned}$ is severely contaminated by foregrounds, up to fully overlap with the power spectrum of \textbf{X}$_R^F$ for $\ell < 50$. Asking GMCA to look for $n_{\rm s}=$ 2 morphologically diverse components is not enough to pinpoint the foregrounds. The leap -- in amplitude and behaviour --  the spectra of \textbf{X}$^{\rm cleaned}$ exhibits when passing to the $n_{\rm s}=3$ scenario is a hint for having reached an optimal $n_{\rm s}$, further validated by the convergence the spectrum of \textbf{X}$^{\rm cleaned}$ shows in the $n_{\rm s}=$ 4 and 5 plots. Looking at the power spectra of projections: increasing further number of components $n_{\rm s} > 3$ helps (marginally) to better characterised the foregrounds (almost imperceptibly in these plots, with the exception of  the galactic free-free component: the blue dotted line keeps decreasing in amplitude with increasing $n_{\rm s}$), but it comes at the expense of having more leakage of the true signal (although imperceptible by eye as well). Setting $n_{\rm s } = 3$ is optimal in this observational set-up, as already proven by Figs.~\ref{fig:Rl_m00} and ~\ref{fig:Rnu_00}, and, noteworthy, we can reach this conclusion by examining the power spectrum of  \textbf{X}$^{\rm cleaned}$ alone, without comparing with the ground truth one.

\subsection{Mimicking RFI}
\label{sec:RFI}

\begin{figure}
	\includegraphics[width=\columnwidth]{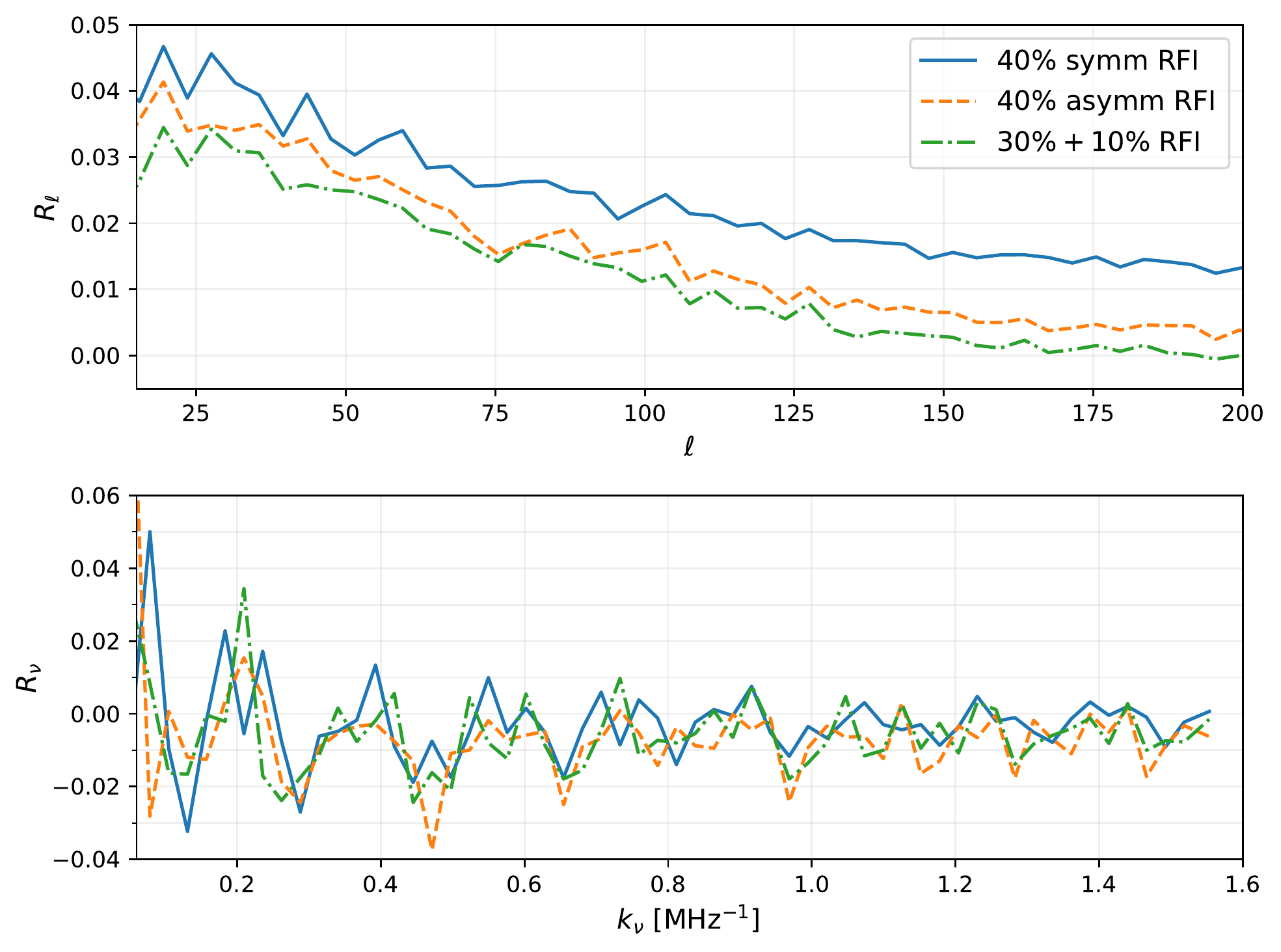}
	\caption{In the top (bottom) panel is the relative difference in angular (radial) power spectrum between the GMCA ($n_{\rm s} =3$) recovered signal and the input cosmological signal. Out of the 200 channels of the simulations, 80 have been discarded: in one chunk at the centre of the frequency interval (solid line), in one chunk in the first half of the frequency interval (dashed) and in two separated chunks (dot-dashed). Frequency-incomplete data does not compromise the GMCA foreground separation.
	}
	\label{fig:gmca_RFI}
\end{figure}

When performing radio observations, whole channels are  discarded  due to irreversible contamination by radio-frequency interference (RFI) generated for instance by FM radios and television stations, cellular network of mobile phones, satellites, and so on. Even in radio-quiet areas designated and protected for those experiments, RFI flagging is usually still necessary. For instance, for the on-going MeerKLASS 21cm intensity mapping L-band preliminary observations, roughly $40 \%$ of the data in two separated chunks of flagged channels is typically discarded.  As previously pointed out, the performance of the foreground removal depends on the number of channels and on the frequency range covered by the data-cube. This motivates the question: how does having missing channels effect the foreground removal? 

We mimic the RFI flagging effect  by removing $40 \%$ of the channels in the simulation and run GMCA on the $60 \%$ that is left, i.e. on 120 channels in our case. We adopt three flagging scenarios, removing channels: (i) in one chunk at the centre of the frequency interval, (ii) in one chunk at the beginning of the frequency interval (remaining with the first $10\%$ of channels and the last $50\%$) and (iii) in two chunks of different lengths (in the order: $20\%$ good, $30\%$ flagged, $20\%$ good, $10\%$ flagged, $20\%$ good). Results are shown in Fig.~\ref{fig:gmca_RFI}, in terms of recovery of angular scale information $R_\ell$ in the top panel and of parallel scale information $R_\nu$ in the bottom panel; the different line styles correspond to the three RFI scenarios. GMCA has been run setting $n_{\rm s} = 3$. The overall bias level in the angular power spectra of the cleaned maps is analogous with what we measure for a non-RFI-contaminated data-cube composed by 120 channels (i.e. a situation between the left and middle panel of Fig.~\ref{fig:Rl_m00}). Also the scale dependence of $R_\ell$ is not stronger than that of the continuous data-cube case. Among the three different RFI scenarios, the last one with three frequency-discontinuous chunks of data is slightly better performing, probably due to the better frequency-coverage of the data. Also the radial power spectrum results $R_\nu$  in the bottom panel of Fig.~\ref{fig:gmca_RFI}  are consistent with those obtained with continuous data-cubes in Fig.~\ref{fig:Rnu_00}, being below $5\%$ at large scales and going down to $\approx 1\%$ for small scales, with essentially no difference among the three RFI scenarios. When run on RFI-affected data-cubes, GMCA yields to mixing matrices $\tilde{A}$ with {\it jumps} in columns, thus recognizing the discontinuous  nature of the data and being able to benefit from the whole data available without the need to partition and lose frequency information of the components.

 Summarising, it is reliable and still effective to use GMCA with flagged -- i.e. discontinuous -- data.

\subsection{Masking}
\label{sec:mask}

\begin{figure}
	\includegraphics[width=\columnwidth]{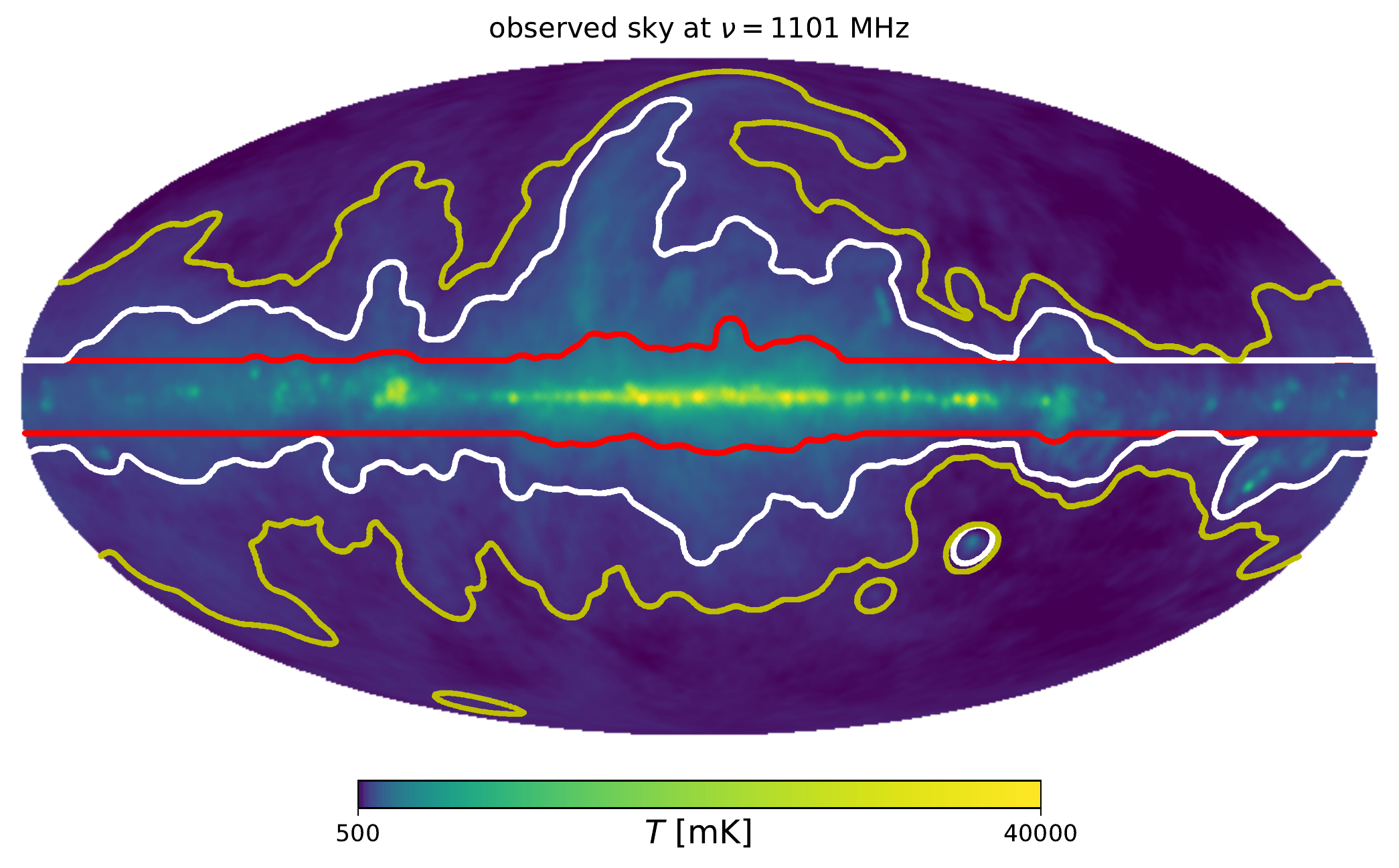}
	\caption{Total temperature map of channel $\nu \in [1100 - 1102]$ of the simulation. We over-plot with solid coloured lines the different masks we use, covering the brightest pixels up to the $10$, $25$ and $50\%$ of the sky. As expected, it is the galactic plane to be masked out, up to the synchrotron North Polar Spur for larger masks.
	}
	\label{fig:masks}
\end{figure}

\begin{figure}
	\includegraphics[width=\columnwidth]{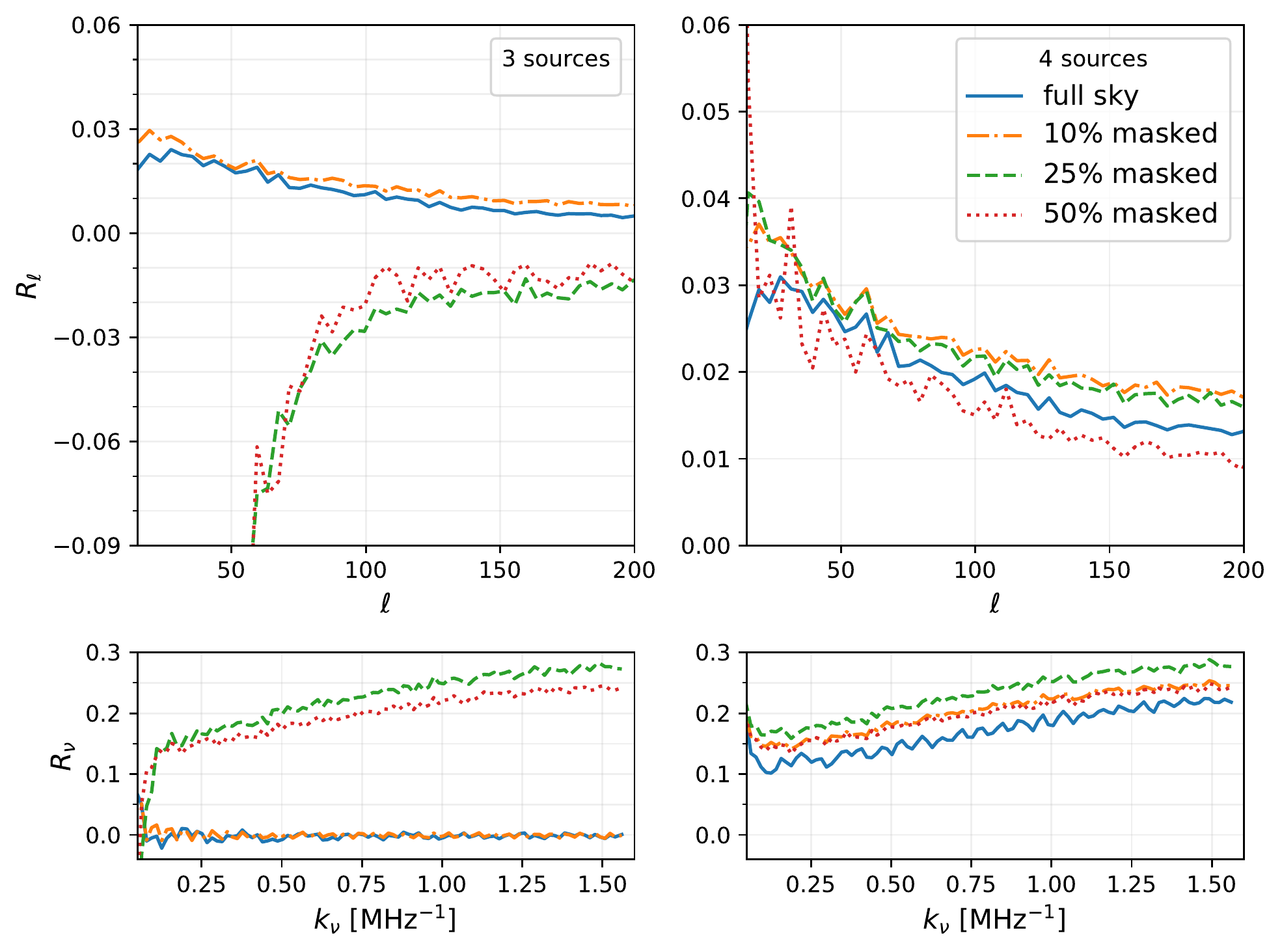}
	\caption{Relative difference in power spectrum between the GMCA recovered signal and the input cosmological signal and instrumental noise: angular $R_{\ell}$ (radial $R_{\nu}$) in the top (bottom) row. GMCA has been run on the data-cube without polarization leakage and using $n_{\rm} =$ 3 (4) number of sources in the left (right) column, and masking out the $10\%$ (dotted line), $25\%$ (dash-dotted), $50\%$ (solid) or using the full-sky maps (dashed). Masking out the region where galactic synchrotron and free-free emissions are more intense, makes it harder for GMCA to reconstruct them. Increasing the number of sources can overcome this at the angular power spectrum level (top right), but the radial one is nevertheless compromised (bottom panels). We note that the masking has an effect on the  power spectrum estimation, for the angular one it reduces the number of large modes available (but this affects mainly scales $\ell < 10$) for the radial it reduces the number of lines-of-sight available.
	}
	\label{fig:res_mask}
\end{figure}

It has been reported that masking the angular regions where foregrounds are more intense benefits the foreground cleaning process \citep{wolz2014,alonso2015,bigot-sazy2015,olivari2016}. We test if this is also the case for the cleaning performed with GMCA, masking out the pixels of the sky where the simulated observed temperature is brightest. We consider brightness thresholds that lead to masks covering the 10, 25 and $50\%$ of the full-sky, inevitably hiding the galactic plane, as shown in Fig.~\ref{fig:masks}.

The wider the mask, the less the pixels and the information GMCA relies on, making unfair a direct comparison of the exercise of this Section with the previous ones. Nevertheless, it can tell us whether covering the most-contaminated region helps the cleaning in the leftover area. 

Our findings are summarised in Fig.~\ref{fig:res_mask}: in the top row the angular power spectrum relative difference $R_{\ell}$, in the bottom the radial counterpart $R_{\nu}$, for runs of GMCA looking for $n_{\rm s} =$ 3 (left) and 4 sources (right column). In the  $n_{\rm s} =3$ scenario, GMCA struggles more to identify the foregrounds in the masked data. In the case of masks of $25\%$ and $50\%$, $R_{\ell}$ is negative, thus the spectrum of the cleaned maps is higher than that  of the ground truth: we can push the number of sources to look for, as we do in the right panel. For  $n_{\rm s} =4$, results for the masked scenarios are indeed closer to the full-sky reference, expect at the very large-scales where anyway the angular power spectrum estimation is affected by having less large-modes at disposal due to the partial-sky maps.
On the other hand, looking at $R_{\nu}$ in the lower panels, the $10\%$-mask does not compromise the recovery of information in the radial direction for $n_{\rm s}=3$, and increasing to $n_{\rm s}=4$ does not improve the $R_{\nu}$ level for the $25\%$ and $50\%$ masked cases.

Masking the most contaminated pixels does not help the GMCA reconstruction. On the contrary, we suspect that the morphological detection part of the algorithm (sparsity in the wavelet domain) characterises contaminants better when their features are strongly present. This is at odds with other foreground removal methods and yields to (i) the advantage of  working with the full data-set available and (ii) to more flexibility with the choice of the survey target sky area to begin with, allowing for survey designs with greater commensality with other science scopes (e.g. galactic astrophysics).

We stress again that the masking under study in this Section refers to an a-posteriori covering of bright pixels in the available data. Real surveys do have a mask -- footprint --  on their own, as it is highly improbable to observe the full-sky. The study of the performance of GMCA in different regions of the sky -- with different levels of sparsity of the foregrounds -- is another issue that merits more detailed work.

\subsection{Including the polarization leakage}
\label{sec:pol_leak}

\begin{figure}
	\includegraphics[width=\columnwidth]{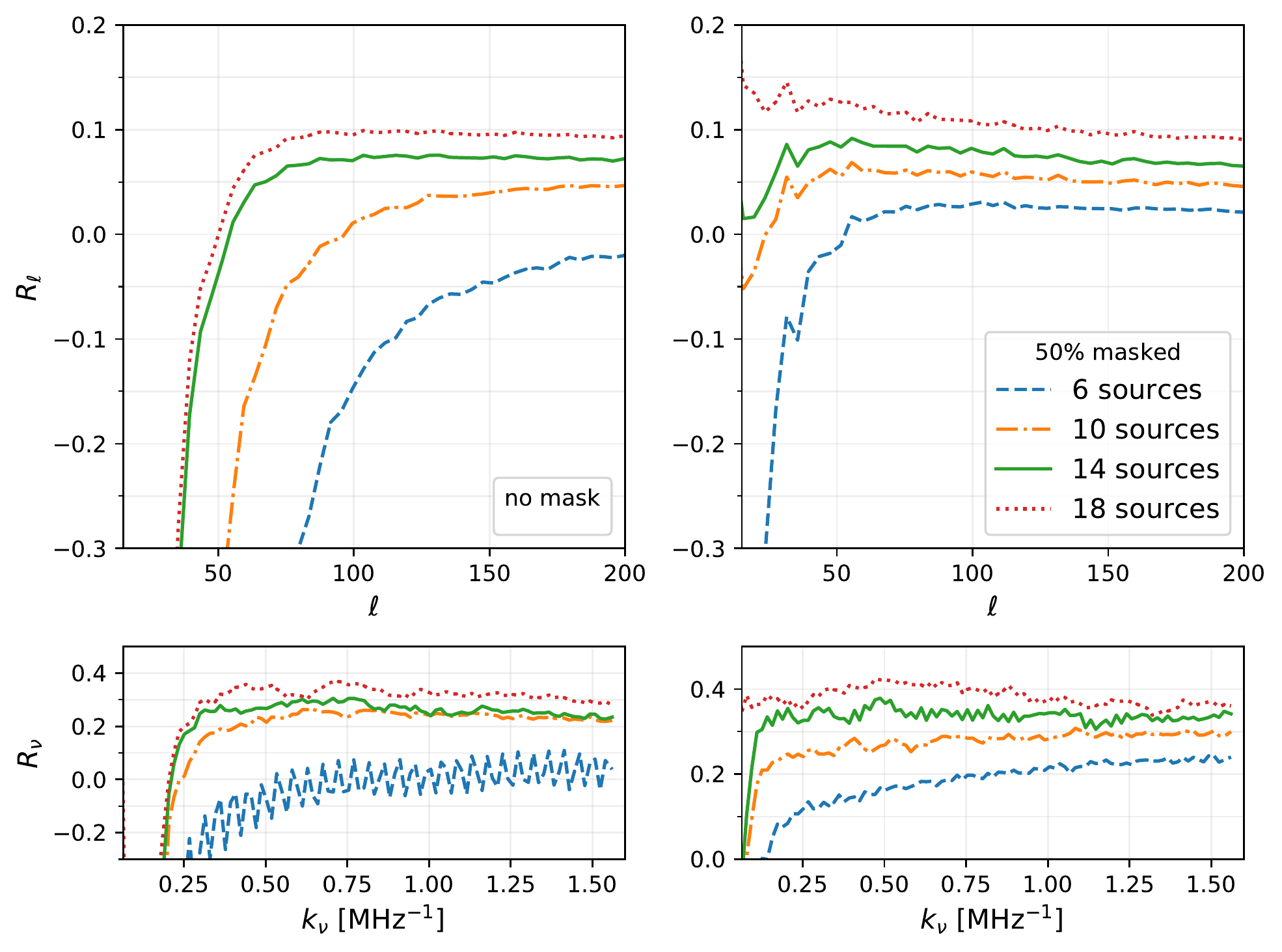}
	\caption{Relative difference in power spectrum between the GMCA recovered signal and the input cosmological signal and instrumental noise: angular $R_{\ell}$ (radial $R_{\nu}$) in the top (bottom) row. GMCA has been run on the data-cube that also contains a $0.5\%$ polarization leaked into the unpolarized signal.	On the left panel, we consider the full-sky maps, on the right $50\%$ of the maps has been masked out. Different line styles and colours correspond to different number of components $n_{\rm s} =6$, 10, 14 and 18 GMCA has been run with. When excluding the galactic plane region (right panel),  GMCA reconstructions improve. Anyway, also for the full-sky scenarios in left panel, results are encouraging at small scales.
	}
	\label{fig:gmca_PL}
\end{figure}

\begin{figure}
	\includegraphics[width=\columnwidth]{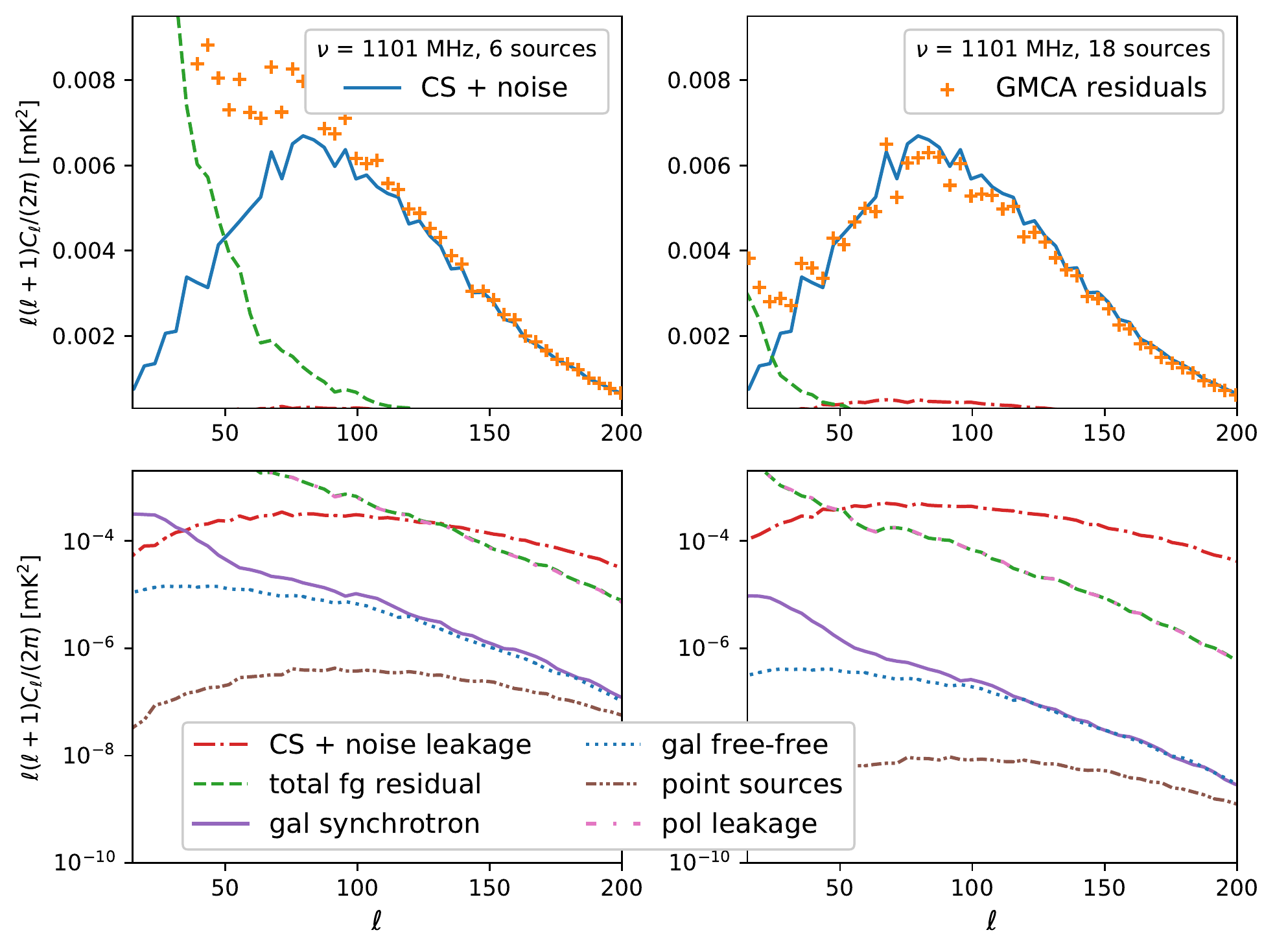}
	\caption{GMCA results for the  $\nu \in [1096 - 1098]$~MHz channel of the simulation full-sky and with polarization leakage.  In the top panels are the angular power spectra of the input cosmological signal and noise  $\textbf{C} + \textbf{N}$ (solid line) and the GMCA recovered signal \textbf{X}$^{\rm cleaned}$ (plus signs). In the bottom panels are the leakage of the input cosmological signal and noise into the GMCA found foregrounds \textbf{X}$_F^{CN}$ (dash-dotted line), and the residuals of the input foregrounds left over in the GMCA recovered signal\textbf{X}$_R^F$  (dashed line) and of the individual foreground component (see legend). The left (right) column corresponds to a GMCA run with $n_{\rm s} = 6$ (18) number of components. polarization leakage (pink dashed) is the worst-identified foreground as it dominates the total foreground residual budget \textbf{X}$_R^F$. The latter is more under control in the right panel, although there is an increase of the cosmological signal that gets lost: the increase of the \textbf{X}$_F^{CN}$ power spectrum from the left to the right scenario is hardly visible in the bottom panels (with logarithmic scales), but it is evident as we start seeing it in the top right panel too, less than an order of magnitude away from the recovered \textbf{X}$^{\rm cleaned}$ power spectrum.
	}
	\label{fig:Clres_PL_fgs}
\end{figure}

Up to now, we have looked at the performance of GMCA on simulated data which do not include polarization leakage. In this Section, we finally add the distressing component in the game.

Our findings are summarised in Fig.~\ref{fig:gmca_PL}, where we plot $R_\ell$ and $R_\nu$ (top and bottom rows) of the results for full and $50\%$-masked sky scenarios (left and right columns respectively) that GMCA yields when run with $n_{\rm s} =6$, 10, 14 and 18 number of components. The addition of polarization leakage undoubtedly makes source-identification by GMCA more troublesome and the number of components to look for has to increase to reach satisfactory levels of cleaning, as it could already be expected by looking at the principal components of the data frequency covariance matrix in Fig.~\ref{fig:eigenv}. Looking at the right panels of Fig.~\ref{fig:gmca_PL}, the situation remarkably improves when we hide the region of the sky where the polarization leakage has the most complex and uneven frequency behaviour (see Fig.~\ref{fig:PL_LoS}). For the left panels case, we can nevertheless make use of the GMCA reconstruction for $\ell > 80$ and $k_\nu > 0.3$ MHz$^{-1}$ for the higher $n_{\rm s}$ considered, as the bias introduced in the recovered power spectrum is scale-independent and, therefore, can be easily taken into account \citep{Cunnington2020} and even marginalised over in cross-correlation analysis.

More about the scale-independence of both $R_\ell$ and $R_\nu$: we can push it to hold for lower scales by increasing $n_{\rm s}$ at the expense of increasing the amplitude of $R_\ell$ and $R_\nu$, i.e. yielding to a $C_\ell^{\rm cleaned}$ and a $P(k_\nu)^{\rm cleaned}$ that underestimate the true spectra. We have a hint about this when looking at the the principal eigenvalues of Fig.~\ref{fig:eigenv}: when polarization leakage is included, there is not a clear discrepancy between foreground eigenvalues and cosmological ones as the transition between the two is smoother, the modes are mixed. Therefore, the risk of increasing $n_{\rm s}$ is to lose progressively more true cosmological signal that leaks in the identified foregrounds \textbf{X}$^{\rm GMCA}$. We illustrate this last point in Fig.~\ref{fig:Clres_PL_fgs}, where we plot results for a single channel for two GMCA runs: with $n_{\rm s} = 6$ on the left column and  $n_{\rm s} = 18$  on the right. The polarization leakage is the least identified of the foregrounds:  it dominates the whole foreground residual (in the bottom panels its pink dashed line $C_\ell$ completely overlaps the green dashed of the total foreground residual). The recovered 21\,cm signal of the $n_{\rm s} = 6$ case (crosses in top left panel) has an angular power spectrum already off at $\ell \simeq 120$ because of the polarization leakage and (marginally) of the galactic synchrotron left in the residuals maps -- further confirmation of the mode mixing. Results are more sound  for the $n_{\rm s} = 18$ case (right panels), although the true 21cm signal that leaks into the detected foregrounds starts becoming relevant: its corresponding red dashed-dotted line enters in the top panel too, where the input signal and GMCA residual live.

Clearly, in this more realistic scenario, we are not anymore able to identify the optimal $n_{\rm s}$ just by looking at the behaviour of the recovered power spectra, as we did in Section \ref{sec:convergence} for the simplified scenario with no leakage. Moreover, we are not assured that by arbitrarily looking for higher numbers of sources $n_{\rm s}$ we have converging results.

Nevertheless, even if the information retained is more compromised when including a polarization leakage component, the resulting bias both in $R_\ell$ and $R_\nu$ is tractable and can be modelled because of its flatness within a range of scales \citep{Cunnington2020}. Overall a compromise has to be looked for, aiming at maximising the foreground identification and minimizing the loss of true signal. This choice should also depend on the scope of the experiment: it is better to overestimates the signal for detecting the 21\,cm emission in cross-correlation with other cosmic tracers, whereas it is important to perform a more aggressive cleaning when aiming for an auto-correlation detection.

We have attempted improving the cleaning in the presence of the polarization leakage, for instance by imposing one column of the mixing matrix\footnote{Setting it equal to the the galactic free-free spectral index, for which there is greater consensus in the community on its expected value at these frequencies \citep{Bennett1992}.} or additionally whitening the data. We do not report any substantial improvements and therefore we choose to not present those results here. We postpone to another study a more in-depth and dedicated analysis aimed at identifying sources that are non-smooth in frequency as the polarization leakage, by using more sophisticated versions of GMCA \citep[e.g. L-GMCA,][]{L-GMCA1, L-GMCA2} or abandoning the full-blind strategy and imposing extra priors, either on the signal or on the contaminants.

\subsection{Comparison with Independent Component Analysis}
\label{sec:fastica}

\begin{figure}
	\includegraphics[width=\columnwidth]{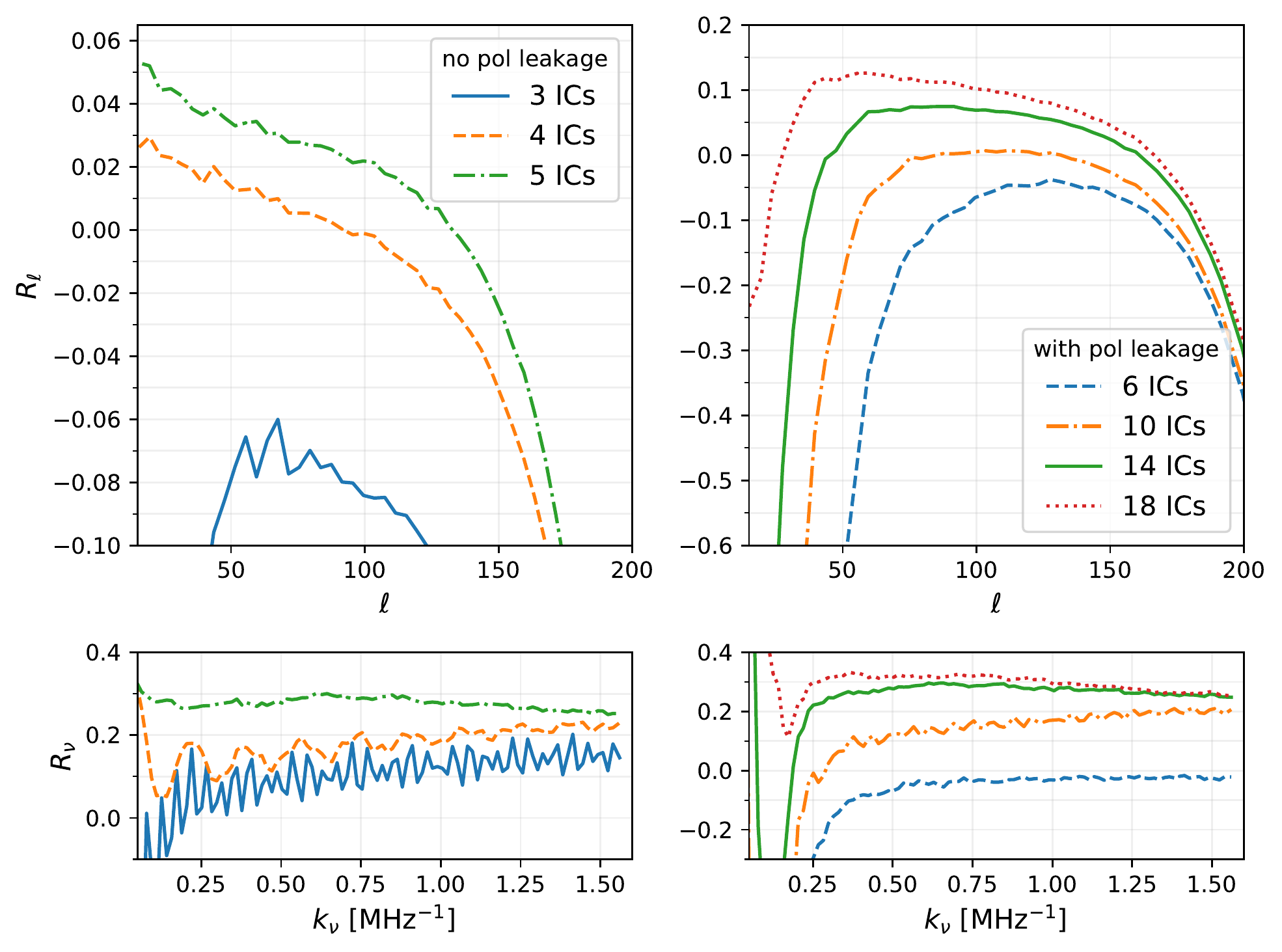}
	\caption{Relative difference in angular and radial power spectrum (top and bottom rows respectively) between the FastICA recovered signal and the input cosmological signal and instrumental noise for a 200 channel full-sky simulated data that (do not) include polarization leakage in the left (right) panels. Different lines correspond to a different number of independent components FastICA has identified. The left (right) panels corresponds to its GMCA counterpart in the left panels of Fig.~\ref{fig:Rl_m00} (right panels of Fig.~\ref{fig:gmca_PL}), same colour-coding.
	}
	\label{fig:fastica}
\end{figure}

\begin{figure}
	\includegraphics[width=\columnwidth]{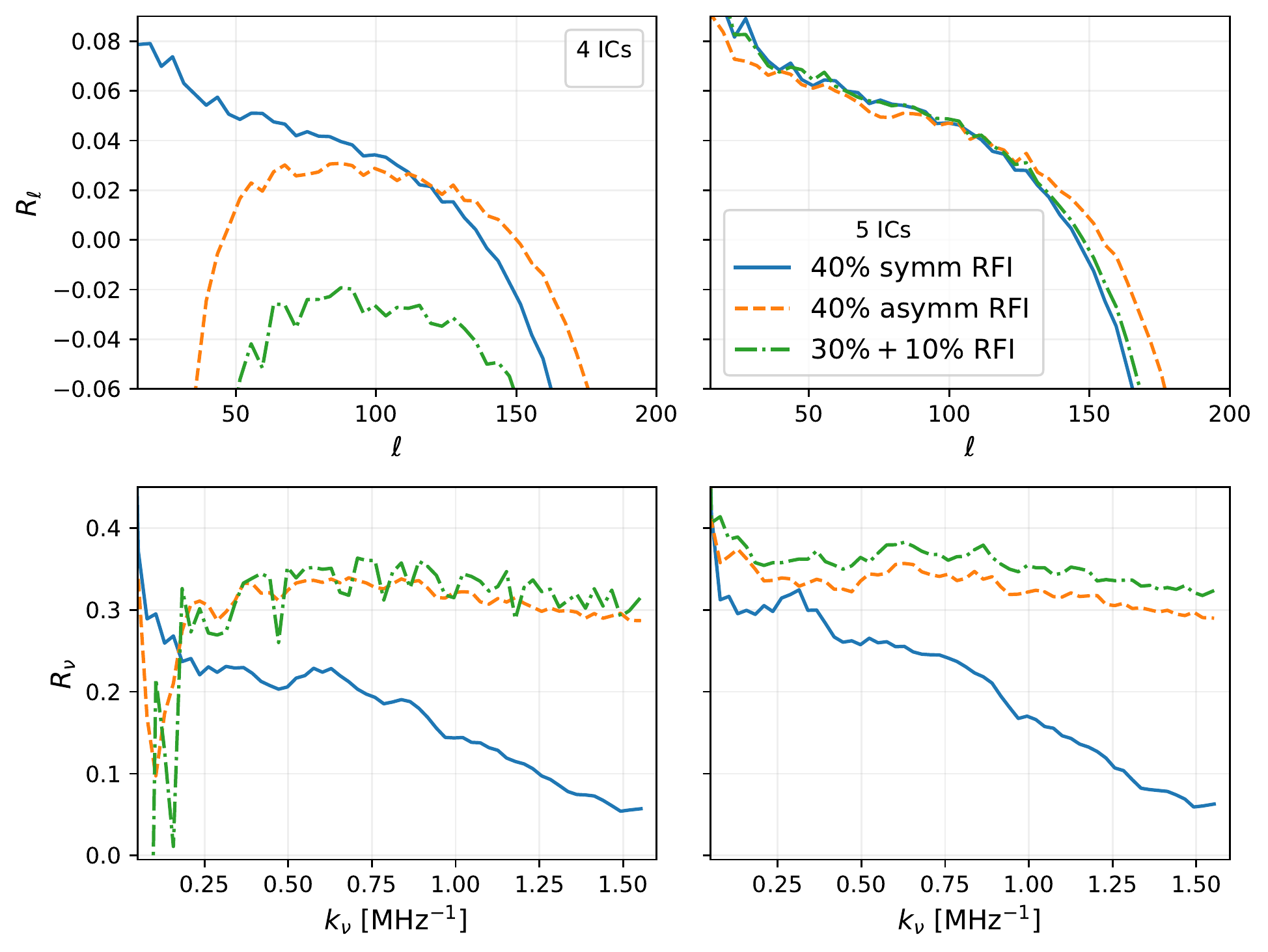}
	\caption{ Relative difference in angular and radial power spectrum (top and bottom rows respectively) between the FastICA recovered signal and the input 21\,cm signal in RFI-compromised scenarios (different line styles). On the left (right) panels, results refer to a FastICA run set to find 4 (5) independent components. Frequency-incomplete data compromises the FastICA foreground separation. Concerning the angular scales (top panels), setting to 5 the number of independent components results in having cleaned maps independent of the RFI-scenario, although the bias in the angular power spectrum is highly scale-dependent. For the information in the radial direction, its recovery does not improve with the increase in number of independent components.
	}
	\label{fig:fastica_RFI}
\end{figure}

For comparison, in this Section we test another foreground cleaning algorithm on the same simulated data. From the currently available and tested methods, we pick the Independent Component Analysis -- in particular the algorithm proposed by \citet{fastica}, FastICA -- that has recently been used on 21\,cm intensity mapping real data by \citet{wolz2017}. In contrast to the GMCA algorithm, which seeks sparse sources in the wavelet domain, the FastICA algorithm looks for statistically independent components by favouring the estimation of non-Gaussian components.

We run FastICA\footnote{\url{scikit-learn.org/stable/modules/generated/sklearn.decomposition.FastICA.html}} on the reference $n_{\rm ch} = 200$ simulation full-sky, with and without the inclusion of polarization leakage. Results are in Fig.~\ref{fig:fastica}, where we plot the relative difference in angular and radial power spectrum, $R_\ell$ and $R_\nu$, between the residuals of the FastICA analysis and the true $\textbf{C} + \textbf{N}$. In the scenario without polarization leakage (left panels), the amplitude of the bias achieved in $R_\ell$ is overall in agreement with that obtained with GMCA (left panel of Fig.~\ref{fig:Rl_m00}), however, the striking difference is the behaviour of $R_\ell$ as function of the angular scale $\ell$. For instance, by setting to 4 the number of independent components (orange dashed line), the resulting average bias in angular power spectrum is of order $\sim3\%$ at large scales, rapidly falls off for increasing $\ell$ and reaches $-12\%$ at $\ell\approx 170$, meaning that FastICA underestimates the true signal at large scales and greatly overestimates it at small scales. We can draw similar conclusions for the scenario with polarization leakage: comparing the right panel of Fig.~\ref{fig:fastica} with the GMCA results in the left panel of Fig.~\ref{fig:gmca_PL}: FastICA reaches similar levels of bias in $C_\ell$ as GMCA, but the relative difference $R_\ell$ has a more complicated angular scale dependence, which makes results harder to interpret and, eventually, foreground cleaning effects harder to model.
Concerning the radial direction, in the scenario with no polarization leakage (bottom left panel of Fig.~\ref{fig:fastica}) FastICA needs 5 independent components to reach a scale independent $R_\nu$, which has amplitude of $30\%$, and with the inclusion of polarization leakage (bottom right), $R_\nu$ displays overall the same levels as for the GMCA reconstructed maps (bottom left of Fig.~\ref{fig:gmca_PL}).

Interestingly, we find a salient difference with respect of GMCA in the RFI-affected scenario. We run FastICA on the same cropped data-cube as described previously in Section \ref{sec:RFI}; results are in Fig.~\ref{fig:fastica_RFI}, with the same colour coding of the GMCA counterpart in Fig.~\ref{fig:gmca_RFI}. Again, the $R_\ell$ quantity is much more scale dependent for the residuals obtained with FastICA. Setting the number of independent components to 4 -- that has been proven optimal in the non-RFI-contaminated case -- gives different $R_\ell$ curves for the different RFI scenarios; setting the components to 5 leads to more consistent results, however, the strong dependence on angular scale is still present. Concerning the radial direction, FastICA yields to  $R_\nu$ that are higher ($\approx 35\%$) than what obtained with GMCA (few percent); moreover, for the symmetric RFI scenario, $R_\nu$ is scale-dependent even when increasing the number of independent components to 5.

\section{Conclusions and perspectives}
\label{sec:conclusions}

The purpose of this work is investigating the foreground cleaning of 21\,cm intensity mapping data performed with the GMCA algorithm, assessing how much information we can recover in terms of the 21\,cm field power spectrum. We use a full-sky simulation of the sky in the $900 - 1400$ MHz frequency range composed by the 21\,cm signal, the expected astrophysical foregrounds, a polarization leakage component, the smoothing due to the telescope beam and the thermal noise of the instrument.  We hereby summarise our main findings. 

\begin{enumerate}
	\item When polarization leakage is not included, we find $n_{\rm s} = 3$ components appropriate for the GMCA cleaning, leading to residuals that underestimate the ground truth angular power spectrum by $\lesssim 2\%$  (channel average) and reproduce at sub-percent level the radial power spectrum for $k_\parallel \gtrsim 0.02\,h$ Mpc$^{-1}$.
	\item When we increase the complexity of the simulation, higher number of sources $n_{\rm s}$ is needed, and results convergence with increasing $n_{\rm s}$ is not assured.
	\item Including polarization leakage and adopting $n_{\rm s} = 14$ sources, the angular power spectrum is recovered with a scale-independent $\approx 7\%$ bias for scales $\ell> 75$ and the radial counterpart with a scale-independent $20 - 30 \%$ bias for scales $k_\parallel \gtrsim 0.1\,h$ Mpc$^{-1}$.
	\item The latter biases improve if we mask the sky region where the adopted polarization leakage component has the most fluctuating behaviour in frequency.
	\item The GMCA source separation benefits from using the highest number of channels available. That is to say, for a fix band-width of the experiment, it has to be privileged the thinnest binning possible.
	\item The GMCA cleaning benefits when it runs on the available data for the full range in frequency, rather than partitioning the data in smaller chunks.
	\item The latter  still holds for incomplete data-cubes, i.e. GMCA performance does not deteriorate for RFI-contaminated data.  
	\item The GMCA source separation does not benefit from masking the sky regions where foregrounds are stronger. For instance, the foreground removal is not less successful in the galactic plane region.
\end{enumerate}

The latter point implies that no data is wasted a-posteriori and that, at the planning stage, experiments do not need to take into account foreground-avoidance for designing the survey footprint, letting focus be rather on issues as overlaps with other samples for cross-correlation and validation purposes, commensality and so on.

As said, when dealing with polarization leakage, cleaning improves when knowing and masking the pixels where this component has a fluctuating temperature contribution as function of frequency. However, this result depends on the model we have adopted for the leakage. Work is needed for a more physical-motivated polarization leakage model, built upon more recent diffuse polarized emission data and galactic magnetic field structure data (e.g. extending the work by \citet{Spinelli2018} to the frequencies of interest).

To our knowledge, this is the first work that studies the possibility of a blind removal for a troublesome foreground component as the polarization leakage. More is still to be done and many are the perspectives of this work. We plan to keep adapting GMCA for better dealing with the leakage and also with other sources of systematics that we did not tackle in the present work, as for instance a more realistic telescope beam that generates mode-mixing in the data and satellite contamination.

For comparison, we also run the FastICA algorithm on the same data-cubes. We can appreciate that -- with respect to FastICA -- GMCA provides results overall more consistent in scale-independence and handles RFI-contaminated data better. More exhaustive comparisons are beyond the scope of this paper. 
GMCA has already been compared with a Gaussian Process Regression method on EoR-like data by \citet{Mertens2018}, who applied GMCA in the Fourier domain. The key assumption underlying the GMCA algorithm is the sparsity of the components to extract in a given domain and, whether it is for EoR or for $z<6$ science, the foregrounds are smoothly distributed in the Fourier domain and therefore not sparse at all. This is  why in this work we prefer a wavelet-based representation to better model the sought-after foregrounds. In short, applying GMCA in a signal representation where the components to be extracted are not sparse is very likely to be less effective, leading to poor separation results. More comparisons between GMCA and other separation methods have been done in the  Cosmic Microwave Background context; \citet{planck_comparison} offers an exhaustive review.

In this work we have proven that the number of sources needed for the cleaning sharply increases with the complexity of the simulated data, and this holds for any blind foreground removal method that assumes data can be linearly decomposed in a fixed number of components as  in equation~(\ref{eq:TF}), e.g. also for FastICA. It is thus important for the community to start testing cleaning algorithms on the most realistic simulations possible. This conclusion does not come unexpectedly as we are aware that in real data analysis, the number of components that is removed is usually higher than what suggested and quoted in simulation papers (e.g. in the most recent analysis  \citet{wolz2017} use 10 and 20 independent components with ICA on GBT data, \citet{anderson2018} use 10 modes with the SVD method with Parkes data; moreover, in both analysis maps are first re-smoothed to further lower resolutions to mitigate the polarization leakage).
It is timely to assess the different systematics in simulations to understand what is at play in the data collected by radio telescopes and to prepare for next surveys.

In this paper we consider full-sky maps. Ongoing work is dedicated to smaller patches and different sky regions, where we expect different foreground contributions and different levels of sparsity that GMCA can rely on. Also depending on the resolution one works with, the sparsity of foregrounds may not always be an appropriate assumption.

Concerning the beam and the noise choices, in this work we consider a single-dish experiment with characteristics of a radio-telescope like the MeerKAT. Nevertheless our analysis is meaningful for other experimental set-ups, also including interferometry-driven 21\,cm intensity mapping experiment as CHIME\footnote{\url{https://chime-experiment.ca}}, Tianlai\footnote{\url{http://tianlai.bao.ac.cn}}, HIRAX\footnote{\url{https://hirax.ukzn.ac.za}} or the proposed PUMA\footnote{\url{https://www.puma.bnl.gov}}: as the decGMCA version of the algorithm performs deconvolution at the same time as the source separation \citep{decGMCA,remi_decGMCA}, it is possible to work directly with the visibility data. This constitutes another interesting line of work. 

In this paper we did not consider the effects a GMCA cleaning would have on cosmological analysis, as we mainly focused on a comparison at the maps/ data-cubes level; we leave this for future work.

For reproducing the results of this article, we make available demonstration scripts and notebooks\footnote{\url{https://github.com/isab3lla/gmca4im}} together with the main simulated maps\footnote{\url{http://doi.org/10.5281/zenodo.3991818}}.

\section*{Acknowledgements}

IPC thanks Marta Spinelli for insightful discussions, Mario Santos and Jingying Wang for MeerKLASS RFI information. This work is supported by the European Union through the grant LENA (ERC StG no. 678282) within the H2020 Framework Program.

{\it Software} NumPy \citep{numpy}; Healpy \citep{Zonca2019_healpy}; Scikit-learn \citep{scikit-learn}; Hickle \citep{Price2018}; Matplotlib \citep{Matplotlib}.

\section*{Data availability}

The simulated data underlying this article are publicly available in Zenodo, at  \url{http://doi.org/10.5281/zenodo.3991818}.




\bibliographystyle{mnras}
\bibliography{example} 



\bsp	
\label{lastpage}
\end{document}